\documentclass[final,1p,times]{elsarticle}

\usepackage{graphicx}

\usepackage{amssymb}

\journal{Astroparticle Physics}

\begin{document}

\begin{frontmatter}



\title{Upgrading and testing the 3D reconstruction of gamma-ray air showers as observed with an array of Imaging Atmospheric Cherenkov telescopes}

\author[label1]{M.~ Naumann-God\'{o}\corref{cor1}}
\address[label1]{Laboratoire Leprince-Ringuet, \'{E}cole Polytechnique, IN2P3/CNRS, F 91128 Palaiseau Cedex, France}
\ead{naumann-godo@llr.in2p3.fr}

\author[label2]{M.~ Lemoine-Goumard}
\address[label2]{Centre d'\'{E}tudes Nucl\'{e}aires de Bordeaux-Gradignan, Chemin du solarium, BP 120, F 33175 Gradignan Cedex, France}

\author[label1]{B.~ Degrange}

\begin{abstract}
Stereoscopic arrays of Imaging Atmospheric Cherenkov Telescopes
allow to reconstruct gamma-ray-induced showers in 3 dimensions,
which offers several advantages: direct access to the shower
parameters in space and straightforward calorimetric measurement of
the incident energy. In addition, correlations between the different
images of the same shower are taken into account. An analysis method
based on a simple 3D-model of electromagnetic showers was recently
implemented in the framework of the H.E.S.S. experiment. In the
present article, the method is completed by an additional quality
criterion, which reduces the background contamination by a factor of
about 2 in the case of extended sources, while keeping gamma-ray
efficiency at a high level. On the other hand, the dramatic flares
of the blazar PKS~2155-304 in July 2006, which provided H.E.S.S.
data with an almost pure gamma-ray sample, offered the unique
opportunity of a precision test of the 3D-reconstruction method as
well as of the H.E.S.S. simulations used in its calibration. An
agreement at a few percent level is found between data and
simulations for the distributions of all 3D shower parameters.

\end{abstract}


\end{frontmatter}


\section{Introduction}\label{into}
During the last four years, an important gain in sensitivity was
achieved in the very-high-energy domain of gamma-ray astronomy,
resulting in an increase of the number of confirmed sources by a
factor of about ten. This progress was made possible by the new
generation of Imaging Atmospheric Cherenkov Telescopes, and
particularly by stereoscopic arrays of such detectors. Presently and
in the near future, the number of individual telescopes in the
different arrays is limited: 5 in H.E.S.S.-II \cite{punch}, 4 in
CANGAROO-3 \cite{cangaroo}, H.E.S.S.-I \cite{hess} and VERITAS
\cite{veritas} and 2 in MAGIC-II \cite{magic}. In the long term,
much larger arrays are considered in order to improve both the
sensitivity (through the rejection of hadronic showers and the
increase of the effective detection area) and the angular
resolution. The 3D-reconstruction of gamma-ray showers developed by
M.~Lemoine-Goumard et al. \cite{3Dmodel} in the framework of the
H.E.S.S. experiment is particularly well suited to large arrays
since correlations between different stereoscopic views of the same
shower are taken into account. Furthermore, in this method,
gamma-rays are selected on the basis of physical properties: the
rotational symmetry of an electromagnetic shower and the lateral
spread of the corresponding Cherenkov photosphere at shower maximum.
The total number of Cherenkov photons emitted by the shower is then
reconstructed, yielding the gamma-ray energy in a straightforward
way. This 3D-model was successfully applied in some H.E.S.S.
studies, both on point-like sources (e.g. the blazar H2356-309
\cite{h2356}) and on extended sources (e.g. the supernova remnant
RX~J0852.0-4622 \cite{velajr}, also
named Vela Junior). \\

In the present article, we first describe a significant improvement
in the model which has been achieved since the original publication
\cite{3Dmodel}. In the 3D-reconstruction method, the shower
parameters are obtained by a maximum likelihood fit to the charge
contents of the pixels of each camera, with the single constraint of
rotational symmetry\footnote{The effect of the weak geomagnetic
field (20$\mu$T on the H.E.S.S. site, due to the South Atlantic
Anomaly) is taken into account in the simulations and in the
calculation of the performance of the method.}. However, in the
version described in \cite{3Dmodel}, no goodness-of-fit parameters
were used in the further selection of gamma-ray showers. This
conservative strategy, in which only a loose convergence criterion
was required, was due to the fact that light fluctuations at the
pixel level were not perfectly controlled and that correlations
between the contents of different pixels were not taken into
account. Moreover, in order to be reliable, a goodness-of-fit
parameter based on individual pixel charges would require a complete
modelling of the effects of the night-sky background. In
section~\ref{sec-goodness} below, we show that the situation is
simpler at the level of each triggered telescope, and that we can
directly compare the total photo-electron yield per camera predicted
by the model to the actually measured one. Furthermore, in
non-triggered telescopes, the model should predict a modest light
yield for genuine gamma-ray showers. Such comparisons allow to
define self-consistency conditions which should be satisfied for
most electromagnetic showers but not necessarily for hadronic ones.
Applying this additional quality criterion to H.E.S.S. data on an
extended source\footnote{In the case of an extended source, no cut
on the angle between the source nominal position and the shower axis
is applied.}, the hadronic background remaining after using standard
3D-cuts is reduced by a factor of about 2, while gamma-ray
efficiency is kept at a high level ($\sim 80$\%). The significance
of a weak gamma-ray signal can thus
be improved by a factor 1.24 with respect to the standard 3D-analysis.\\

The performance of the 3D-analysis in the H.E.S.S. experiment was
estimated from gamma-ray simulations which were also used to
calibrate the energy measurement in this method. The simulation
procedure used in this study consists of two parts: (a) development
of gamma-ray-induced showers in the atmosphere (program KASCADE
\cite{kascade}) and (b) complete description of the detection chain
(program SMASH \cite{smash}), namely optics, triggering electronics
and readout electronics of the H.E.S.S. telescopes. In step (b) the
average evolution of the detector with time is taken into account,
since the efficiency of the whole acquisition chain (mirror
reflectivity, light collectors efficiency, quantum efficiency of the
phototubes, electronic charge measurement) is regularly monitored on
the basis of ring-like images produced by muons \cite{vacanti}
\cite{calib} when telescopes are triggered individually. The
dramatic flares of the blazar PKS~2155-304 in July 2006 \cite{pks06}
in which the average intensity of the source was about 7 times that
of the Crab nebula, provided us with a gamma-ray beam with a very
small contamination of hadronic background, offering a unique
opportunity of checking both the performance of the 3D-model and the
quality of the whole simulation chain with a high accuracy. The
corresponding tests are described in section \ref{sec-testing}.

\section{Characterising the quality of the likelihood fit}\label{sec-goodness}

The 3D reconstruction method is based on a likelihood fit in which
the shower parameters in space are obtained from the charge contents
(in single photo-electron units, hereafter p.e.) of individual
camera pixels (see formula (1) of reference \cite{3Dmodel}). Images
were previously submitted to a cleaning procedure in order to remove
isolated clusters of pixels with a small charge which likely result
from the night sky background. The likelihood fit was restricted to
those pixels retained by the cleaning procedure and to their
immediate cleared neighbours, as explained in \cite{3Dmodel}. On the
other hand, starting from the fitted shower parameters, it is
possible to calculate the total photo-electron yield $Q_{3D}$
expected from the 3D model in each camera. This quantity is also
calculated for telescopes which were not triggered by the processed
event and for which no image is available, since, in the framework
of the H.E.S.S. trigger and acquisition system \cite{trig}, they are
not read out. It should be emphasised that, for a given {\it
triggered} telescope, the predicted charge $Q_{3D}$, as well as the
measured charge $Q_{\rm real}$ to which it will be compared,
encompass the whole field of view (except for those pixels
invalidated by the calibration) and not only the restricted sample
of pixels used in the fit. Extending the comparison to the full
field of view allows a better rejection of hadronic showers,
generally more patchy than those of electromagnetic ones, and in
which light is spilled over a region
larger than the image core.\\

\subsection{Quality criterion on images from triggered telescopes}
\label{sec-qual}
\begin{figure}
\begin{center}
  \includegraphics[width=0.49\linewidth]{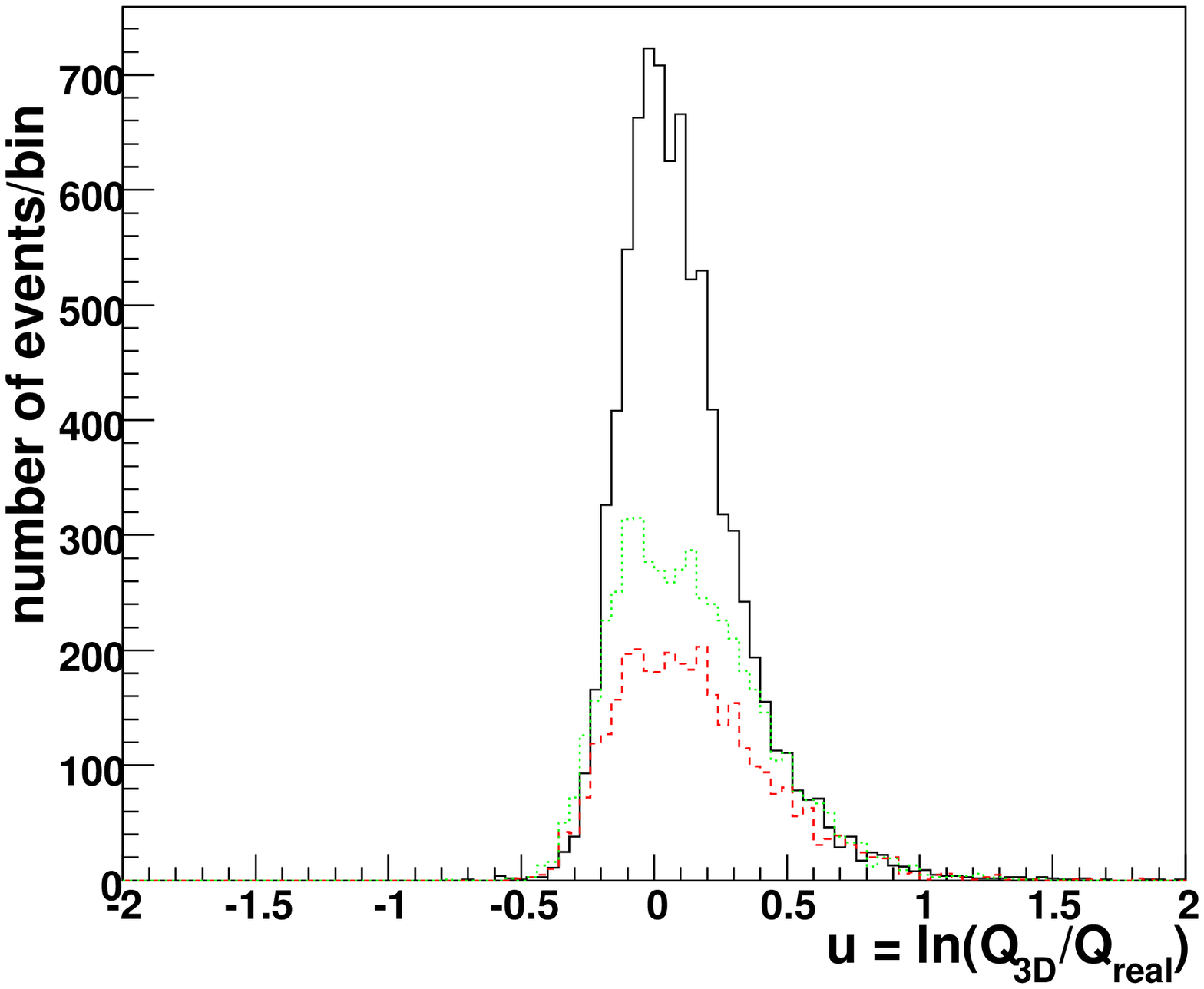}
  \includegraphics[width=0.49\linewidth]{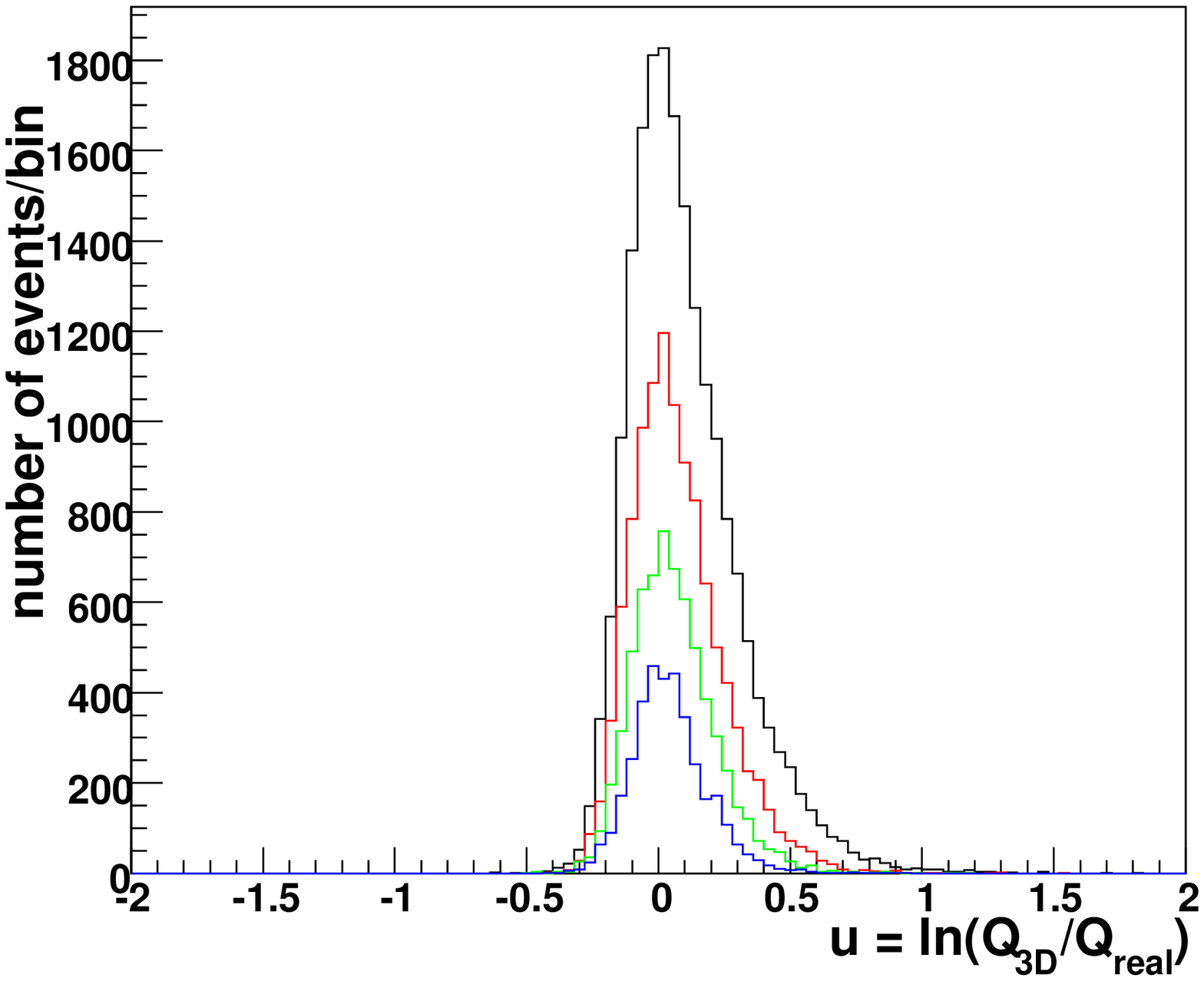}
\caption{\it Left panel: distribution of the variable $u=\ln (Q_{3D}/Q_{\rm real})$ for gamma-ray showers simulated at a zenith angle of $18^\circ$. From top to bottom: showers viewed by 2 telescopes (black), 4 telescopes (green) and 3 telescopes (red). Right panel: distribution of the variable $u=\ln (Q_{3D}/Q_{\rm real})$ for gamma-ray showers simulated at different zenith angles $\zeta$ and viewed by 2 telescopes. From top to bottom: $\zeta= 18^\circ, 37^\circ, 48^\circ$ and $53^\circ$. Gamma-rays were simulated according to a power-law energy spectrum with a spectral index of 2.2.}
\label{fig:u234}
\end{center}
\end{figure}

The compatibility between the model and the image can be
characterised by the variable $u=\ln (Q_{3D}/Q_{\rm real})$.
Simulations of gamma-ray-induced showers at different energies and
zenith angles were used to find the expected $u$ distributions. In
these simulations, the complete detection chain is taken into
account. Simulated images were further submitted to the usual
analysis procedure based on the 3D-model. The distribution of the
variable $u$ is shown in Fig.~\ref{fig:u234} (left panel) for
gamma-ray showers simulated at a zenith angle of $18^\circ$,
according to a power-law energy spectrum with a spectral index of
2.2. Using different simulated spectra (e.g. with spectral indices
greater than 3) does not significantly change the distribution. This
will be further illustrated in section \ref{sec-testing} on real
data from the blazar PKS~2155-304 whose spectral index is 3.3 (see
Fig.~\ref{fig:PKS_gammau2}). The distribution peaks at $u=0$ as
expected and is slightly asymmetric
as shown in Fig.~\ref{fig:u234}, right panel, for events triggering
2 telescopes at different zenith angles. Fig.~\ref{fig:u234} (left
panel) also shows some dependence on the telescope multiplicity,
i.e. the number $n_T$ of triggered telescopes. The difference
between the case $n_T=2$ and that of higher multiplicities is
related to the property of rotational symmetry of the shower which
is assumed but not really checked if $n_T=2$, whereas it acts as an
additional constraint for $n_T \ge 3$, resulting in a slightly
broader distribution of the
variable $u$ in the latter case. \\

\begin{figure}
\begin{center}
\includegraphics[width=0.32\linewidth]{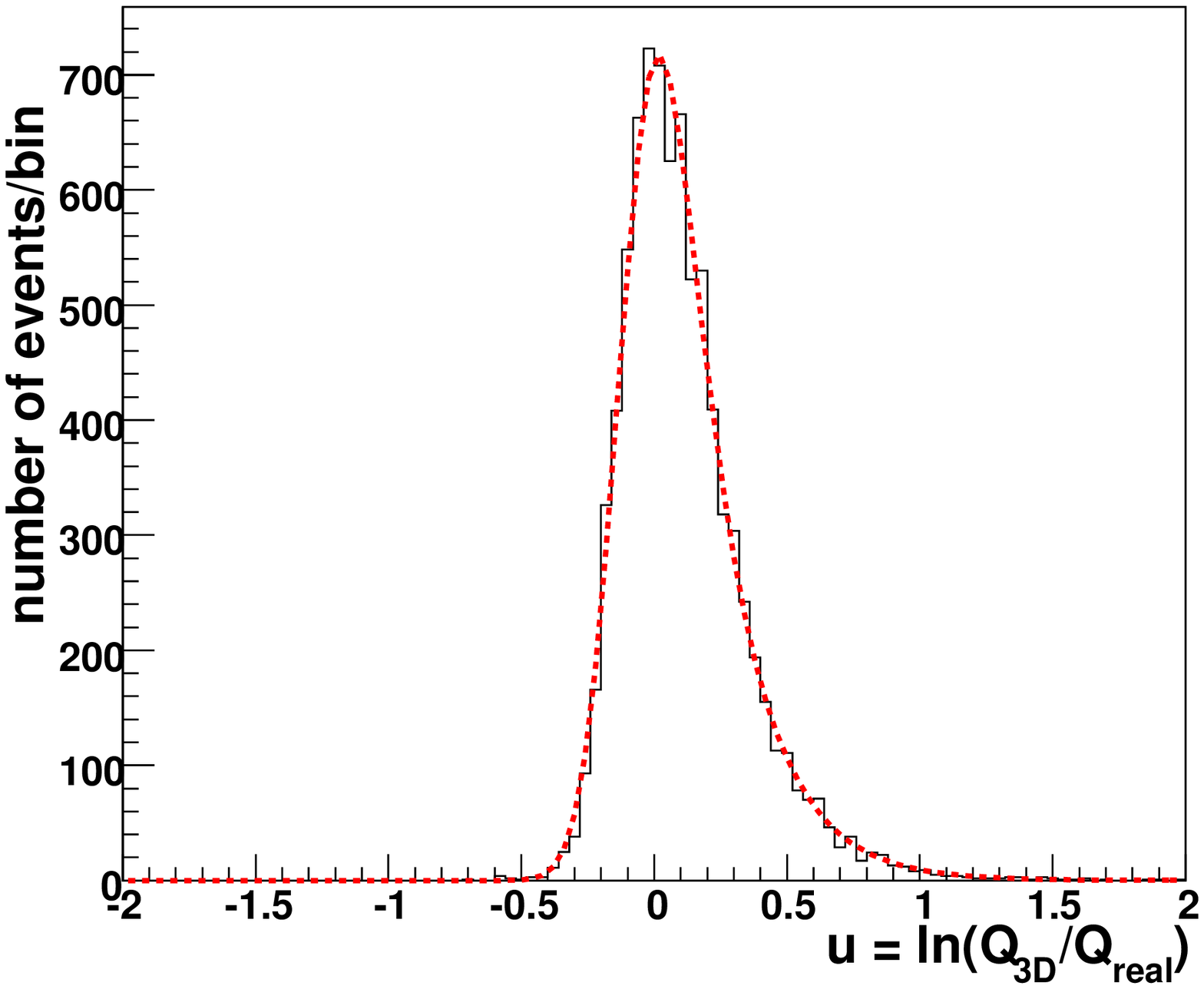}
\includegraphics[width=0.32\linewidth]{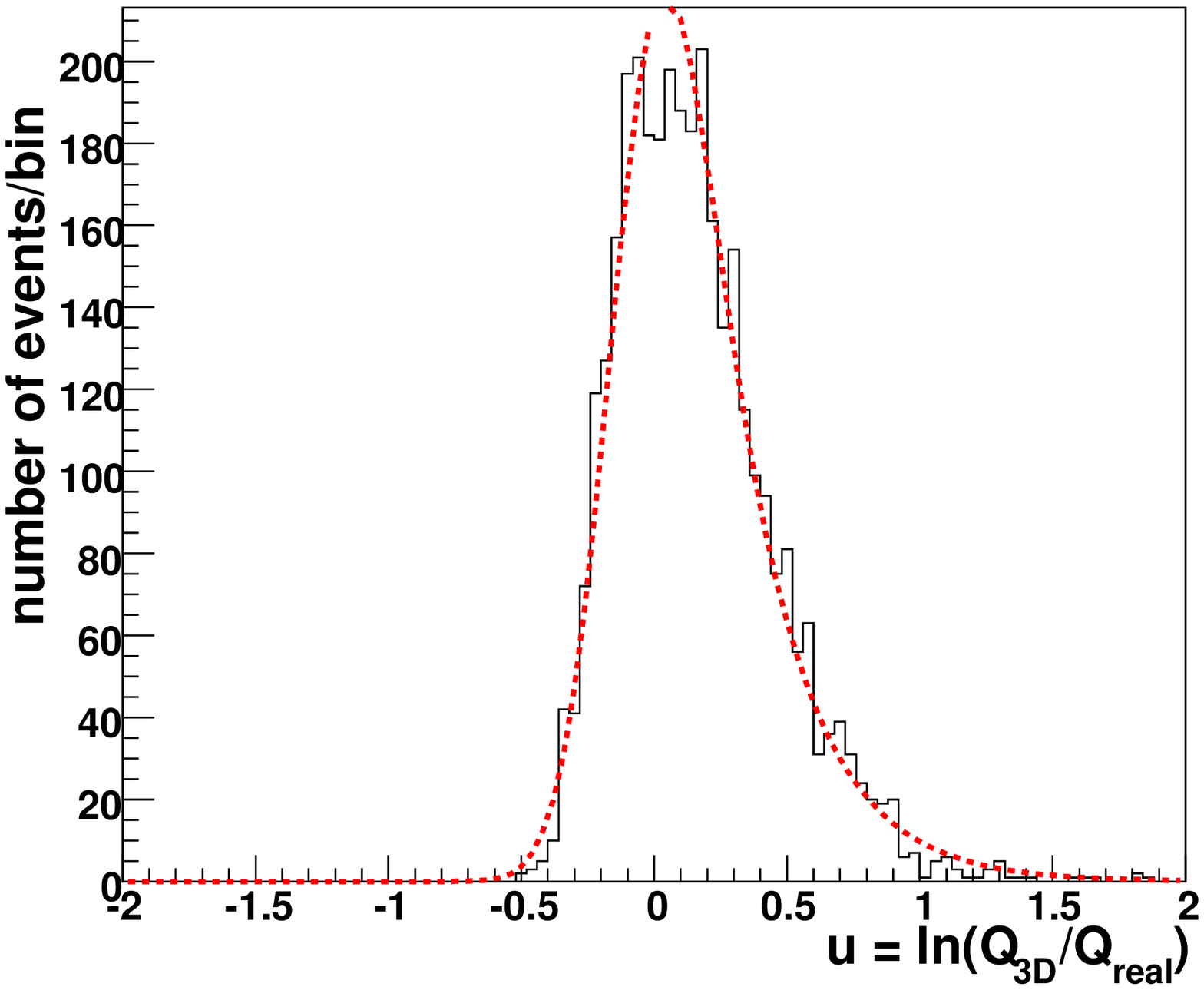}
\includegraphics[width=0.32\linewidth]{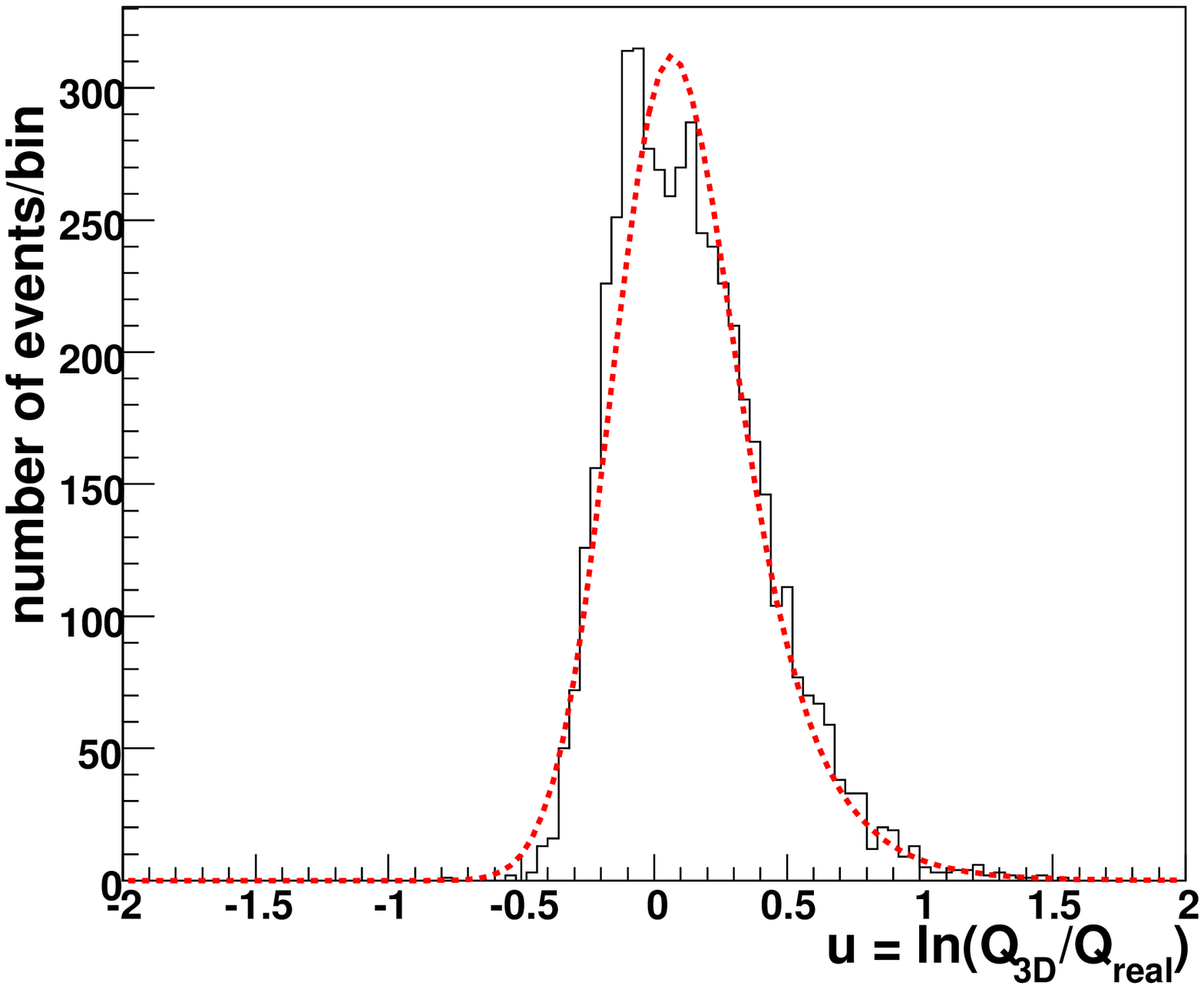}
\caption{\it Distributions of the variable $u=\ln (Q_{3D}/Q_{\rm real})$
for gamma-ray showers simulated at a zenith angle of $18^\circ$~and
viewed by 2, 3 and 4 telescopes (from left to right). Dotted lines show
the distributions fitted according to formula (\ref{eq:param}).} \label{fig:fit234}
\end{center}
\end{figure}

\begin{figure}
\begin{center}
\includegraphics[width=0.97\linewidth]{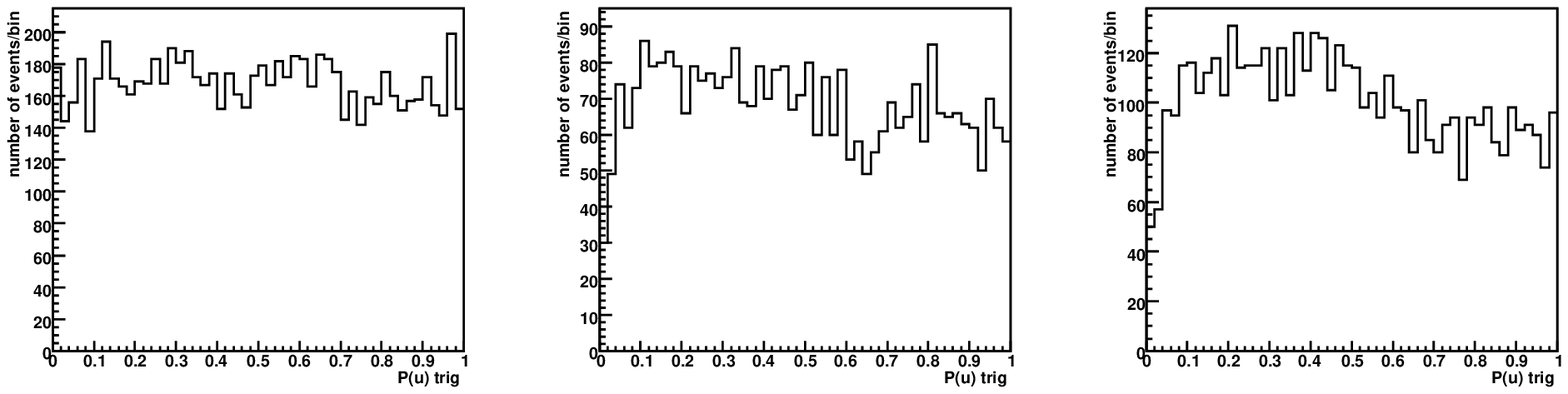}
\includegraphics[width=0.97\linewidth]{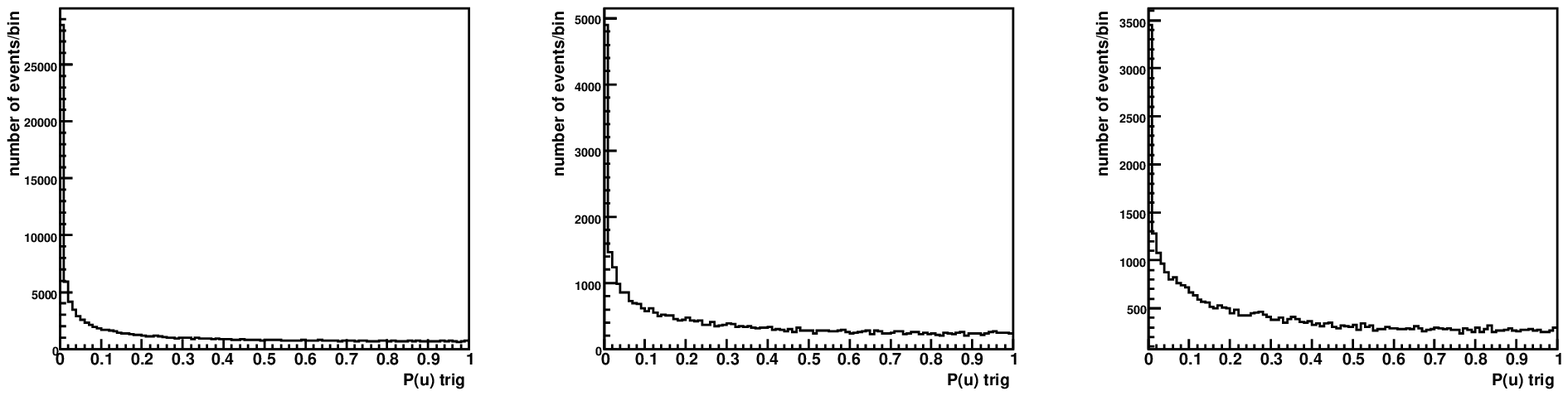}
\caption{\it Distributions of the goodness variable $P(u)$ for
showers viewed by 2, 3 and 4 telescopes (from left to right). Top:
gamma-ray showers simulated at a zenith angle of $18^\circ$,
according to a power-law energy spectrum with a spectral index of
2.2. Bottom: hadronic showers as observed by H.E.S.S. at a zenith
angle of $20^\circ$~in a field of view with no significant gamma-ray
emission.} \label{fig:pi234}
\end{center}
\end{figure}

For $u<0$ the distribution is well described by a Gaussian, whereas
it behaves as an exponential one for large positive values of $u$.
Therefore the probability density function $p(u)$ can be
conveniently described by the convolution of a Gaussian
distribution, with mean value $u_0$ and r.m.s. $s$, and of an
exponential one with mean value $1/\lambda$, namely~:
\begin{equation}
p(u)= K \: e^{-\lambda u} \: {\rm freq}\left( \frac{u-u_0}{s}\right)
\: \: \: \mbox{with} \: \: \: {\rm freq}(x) = \frac{1}{\sqrt{2
\pi}}\int_{-\infty}^x e^{-u^2/2} \, du \label{eq:param}
\end{equation}
the normalization factor $K$ being given by~:
\[ K = \lambda \: \exp(\lambda u_0 -\frac{\lambda ^2 s^2}{2}) \]
The parameters $u_0$, $s$ and $\lambda$ were fitted to the
histograms obtained from simulations. The corresponding fitted
distributions are indicated by the curves superimposed to the
histograms in Fig.~\ref{fig:fit234}. The interest of the analytic
parametrisation (\ref{eq:param}) is that the corresponding
cumulative distribution is easily calculated as~:
\begin{equation}
Q(u) = \int_{-\infty}^{u} p(v) \, dv = {\rm freq}\left( \frac{u-u_0
+ \lambda s^2}{s}\right) - \frac{p(u)}{\lambda} \label{eq:cumul}
\end{equation}
The distribution of $Q(u)$ is thus expected to be uniform for
gamma-ray showers, whereas hadronic ones will exhibit peaks for
small values of $Q(u)$ (corresponding to strongly negative values of
$u$) and for large values (corresponding to strongly positive values
of $u$). It is convenient to fold this distribution in order to
obtain a variable whose behaviour is the same as a $\chi^2$
probability. This is done by using the following variable~:
\begin{equation}
 P(u) = 2 \: {\rm min}(Q(u), 1-Q(u)) \label{eq:cumul2}
\end{equation}
$P(u)$ will be called a ``goodness variable'' in the following. The
distributions of $P(u)$ for those simulated gamma-ray showers
referred to in Fig.~\ref{fig:u234} (left panel) are shown in
Fig.~\ref{fig:pi234} (top panels) and are fairly uniform\footnote{Of
course, these distributions need not be exactly uniform, the effect
of the cuts on $P(u)$ being determined by simulations.}. On the
other hand, distributions of $P(u)$ for hadronic showers detected by
H.E.S.S.~ at a zenith angle of $20^\circ$~in a field of view with no
significant gamma-ray emission are shown in Fig.~\ref{fig:pi234}
(bottom panels) for telescope multiplicities $n_T=2$, 3 and 4 and
show a strong accumulation at
very low values. \\

In practice, the dependence of the parameters $u_0$, $s$ and
$\lambda$ as functions of the telescope multiplicity $n_T$ and the
zenith angle $\zeta$ is determined on the basis of simulations. The
parameter $u_0$ is always found to be compatible with 0 within the
statistical errors, which confirms the quality of the reconstruction
method. As far as the other two parameters $s$ and $\lambda$ are
concerned, the empirical formulae shown in Table~\ref{tab:trigform},
obtained from simulations, give a satisfactory description of the
$u$ distributions in all conditions of observation.

\begin{table}[htbp]
\begin{center}
\begin{tabular}{|c|c|c|}\hline
$n_T$ & $s$  & $\lambda$ \\ \hline 2 & $0.183 - 0.186 \cos \zeta +
0.134 \cos^2 \zeta$ & min(6.3, $9.358 - 4.411\cos \zeta$) \\
3 & $0.084 + 0.087 \cos \zeta$ & $7.141 - 3.528 \cos \zeta$ \\
4 & $0.108 + 0.062 \cos \zeta$ & $8.703 - 4.986 \cos \zeta$ \\
\hline
\end{tabular}
\caption{\it Formulae giving the parameters $s$ and $\lambda$
defined in equation (\ref{eq:param}) as functions of the telescope
multiplicity $n_T$ and of the zenith angle $\zeta$. These formulae
were obtained from simulations.} \label{tab:trigform}
\end{center}
\end{table}


\begin{figure}
\begin{center}
\includegraphics[width=0.7\linewidth]{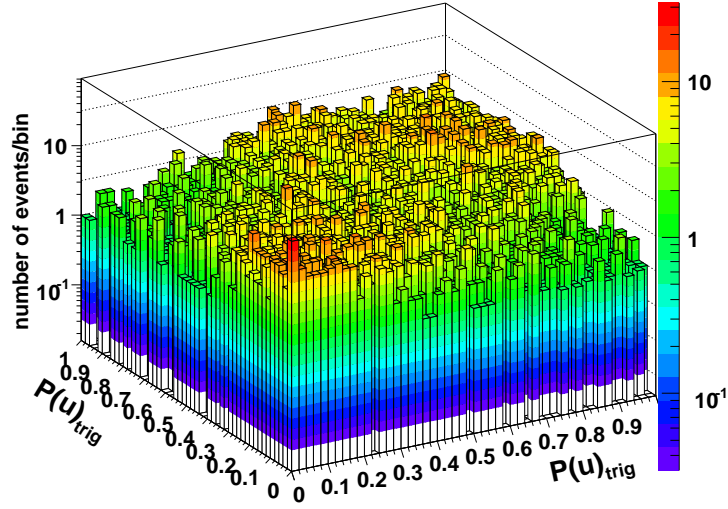}
\caption{\it Gamma-ray showers simulated at a zenith angle of $18^\circ$: the goodness
variables $P(u)$ of the same event viewed by two telescopes are
plotted versus each other.} \label{fig:ggtrig}
\end{center}
\end{figure}

\begin{figure}
\begin{center}
\includegraphics[width=0.7\linewidth]{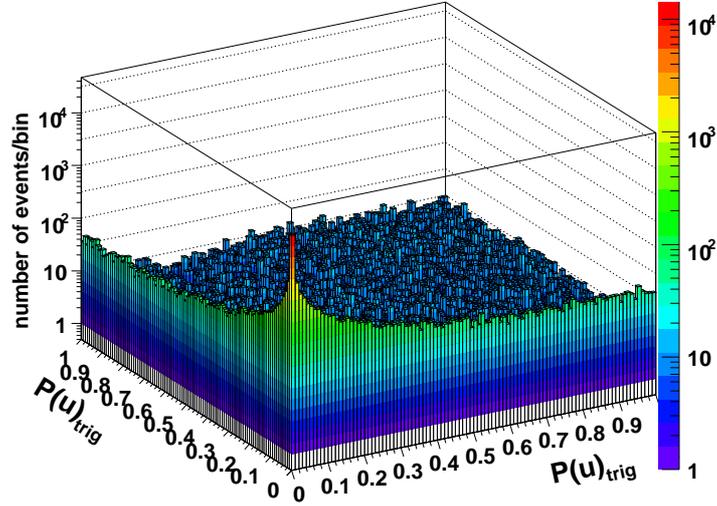}
\caption{\it Hadronic showers as observed by H.E.S.S.~ at a zenith
angle of $20^\circ$~in a field of view with no significant gamma-ray
emission: the goodness variables $P(u)$ of the same event viewed by
two telescopes are plotted versus each other.} \label{fig:hhtrig}
\end{center}
\end{figure}

For each event, formulae (\ref{eq:param}), (\ref{eq:cumul}) and
(\ref{eq:cumul2}) and those given in Table~\ref{tab:trigform} allow
to calculate one goodness variable $P(u)$ per triggered telescope.
For a given event, these quantities are correlated~; in particular,
in the case of genuine gamma-ray showers, there is a clear positive
correlation between the different values of $P(u)$. Therefore, if
$P(u)$ is required to be greater than a minimal value $p_{\rm
min}$ {\it for all triggered telescopes}, the gamma-ray selection
efficiency of such a cut is greater than $p_{\rm min}^{n_T}$. This
is verified from Fig.~\ref{fig:ggtrig} in which two values of
$P(u)$ of each gamma-ray shower are plotted versus each other. The
regions close to the axes are depopulated with respect to that
surrounding the diagonal. The slight excess close to the origin
(about 20 events for a total of 8400 simulated gamma-rays) is
compatible with the contribution of hadron photo-production in the
primary interaction (occurring with a probability of $\approx 2.8
\times 10^{-3}$ \cite{gaisser}), which is taken into account in the
simulation. On the other hand, the same cut provides a powerful
hadronic rejection, as shown in Fig.~\ref{fig:hhtrig} in which two
values of $P(u)$ of each hadronic shower are plotted versus each
other. Those events for which at least one probability is very small
accumulate in two ``walls'' and a spectacular spike is visible at
the origin.

\subsection{Quality criterion on images from non-triggered telescopes}

\begin{figure}
\begin{center}
\includegraphics[width=0.7\linewidth]{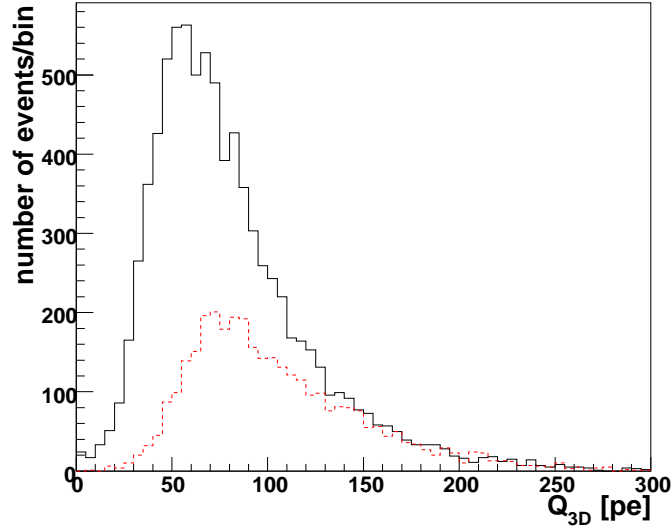}
\caption{\it Distribution of the charge $Q_{3D}$ expected from the
3D-model in non-triggered telescopes for gamma-ray showers simulated
at a zenith angle of $18^\circ$. The solid line corresponds to events
with 2 non-triggered telescopes and the dotted line to events with
only one non-triggered telescope. Gamma-rays were simulated
according to a power-law energy spectrum with a spectral index of
2.2.} \label{fig:qdark}
\end{center}
\end{figure}

\begin{table}[htbp]
\begin{center}
\begin{tabular}{|c|c|c|c|}\hline
$n_T$ & $\ln Q_0$ & $Q_0$ for $\zeta = 20^\circ$  & $\sigma_D$ \\
\hline 2 & $4.367 - 0.092 \cos \zeta $ & 72 & $0.109 + 0.405 \cos
\zeta$
\\  3 & max(4.5, $4.296 + 0.288 \cos \zeta)$ & 96 & $0.184 + 0.259  \cos
\zeta$
\\ \hline
\end{tabular}
\caption{\it Formulae giving $Q_0$ (in p.e.) and the standard
deviation $\sigma_D$ of the distributions of $\ln Q_{3D}$ as
functions of the telescope multiplicity $n_T$ and on the zenith
angle $\zeta$. These formulae were obtained from simulations.}
\label{tab:darkform}
\end{center}
\end{table}

\begin{figure}
\begin{center}
\includegraphics[width=0.46\linewidth]{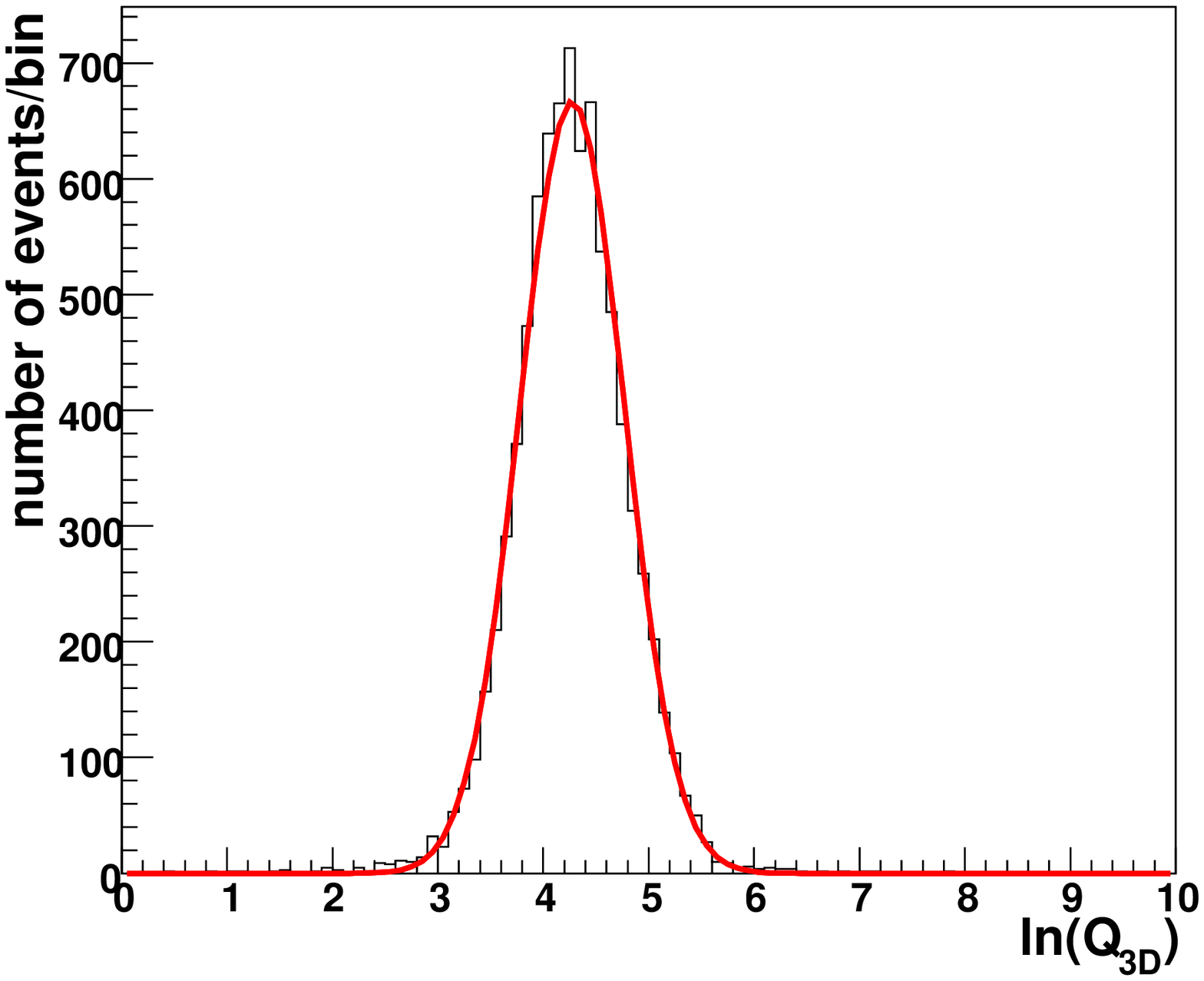}
\includegraphics[width=0.46\linewidth]{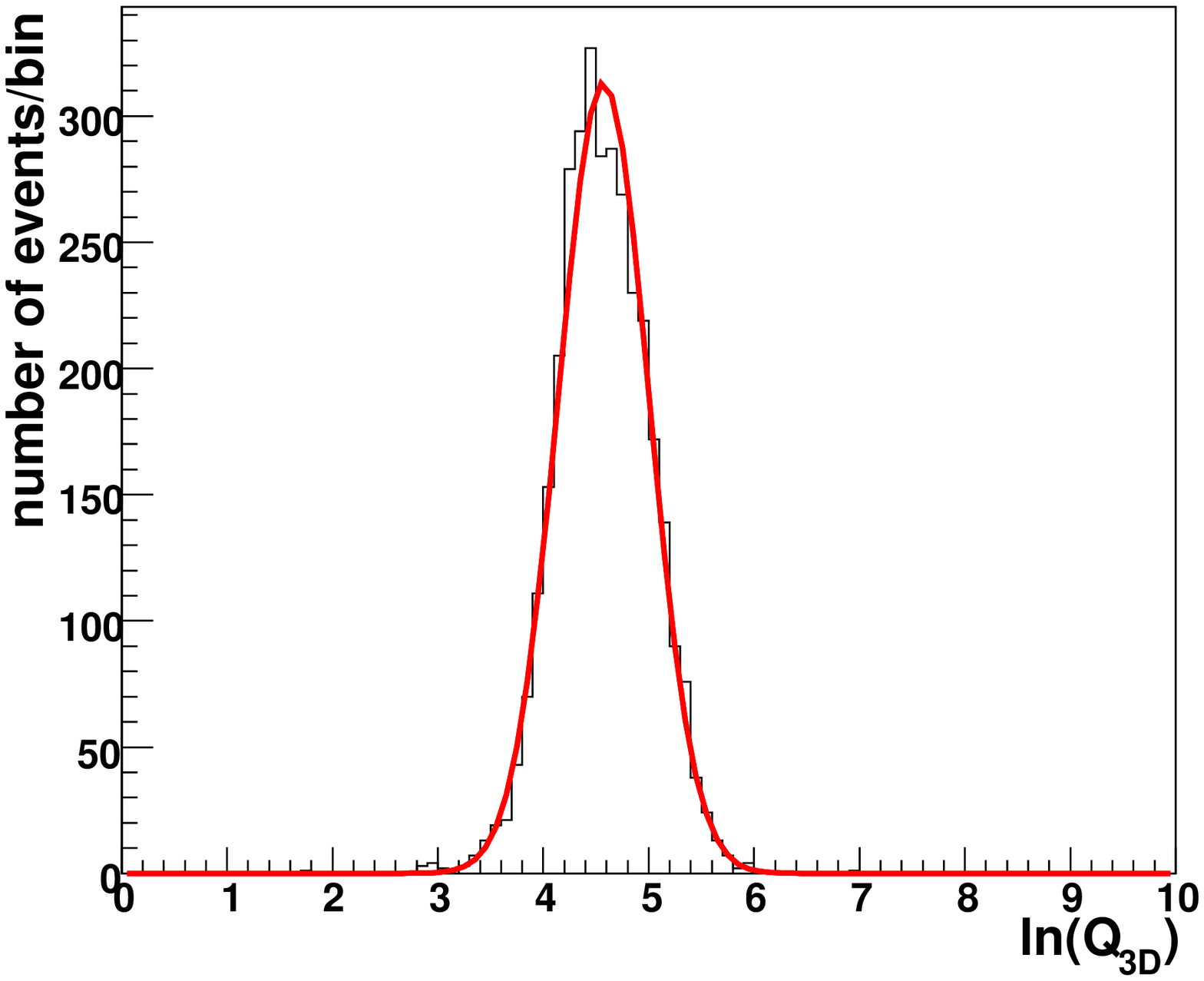}
\caption{\it Distributions of $\ln (Q_{3D})$ for gamma-ray showers
simulated at a zenith angle of $18^\circ$. Left panel: events with 2
non-triggered telescopes ($n_T=2$) and $Q_{3D}>20$~p.e. Right panel:
events with 1 non-triggered telescope ($n_T=3$). Gamma-rays were
simulated according to a power-law energy spectrum with a spectral
index of 2.2.} \label{fig:v23}
\end{center}
\end{figure}

The preceding simulations of gamma-ray showers were also used to
find the distribution of the light yield expected from the 3D-model
in non-triggered telescopes. The same sample of simulated gamma-ray
showers as in the preceding subsection was used to find the
distributions of the charge $Q_{3D}$ predicted by the reconstruction
method in non-triggered telescopes. They are shown in
Fig.~\ref{fig:qdark}, both for $n_T=2$ (i.e. 2 non-triggered
telescopes) and for $n_T=3$ (i.e. one non-triggered telescope). It
can be verified that typical values of this undetected charge range
between 50 and 100~p.e. Furthermore, it turns out that
the distribution of $\ln (Q_{3D})$ is practically Gaussian (see
Fig.~\ref{fig:v23}).

\begin{figure}
\begin{center}
\includegraphics[width=0.95\linewidth]{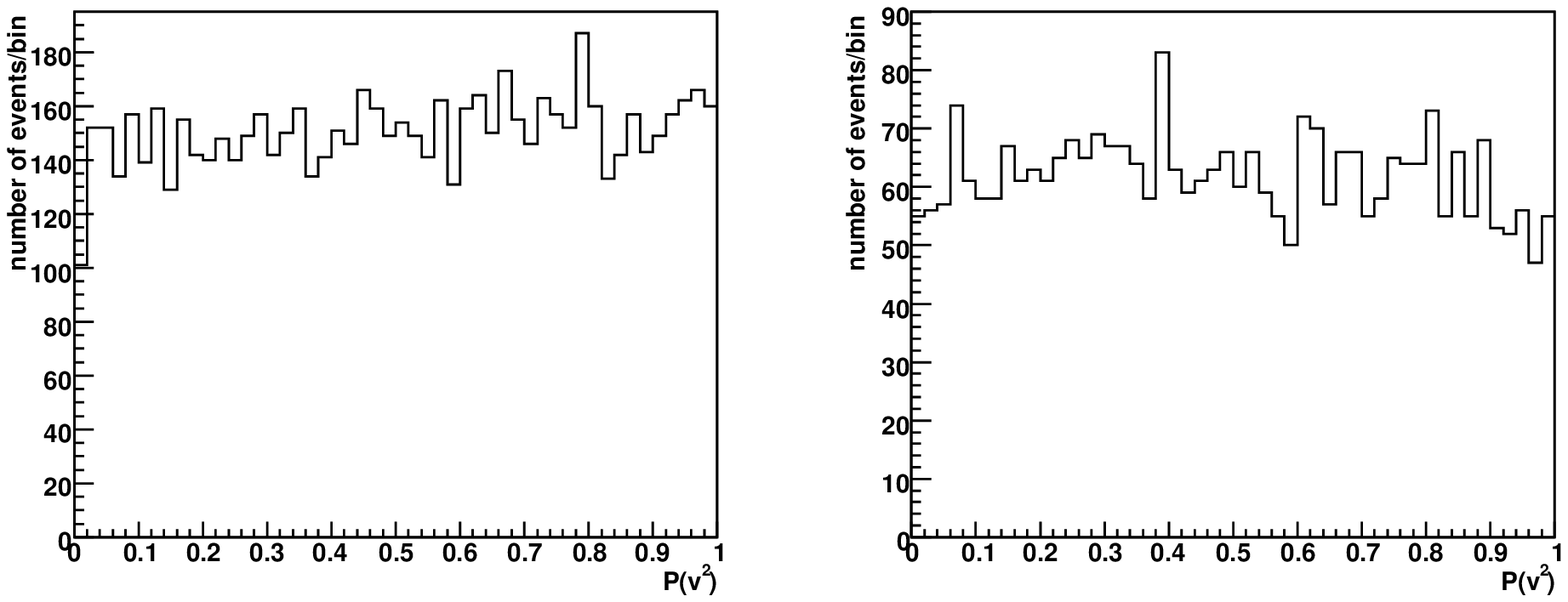}
\includegraphics[width=0.95\linewidth]{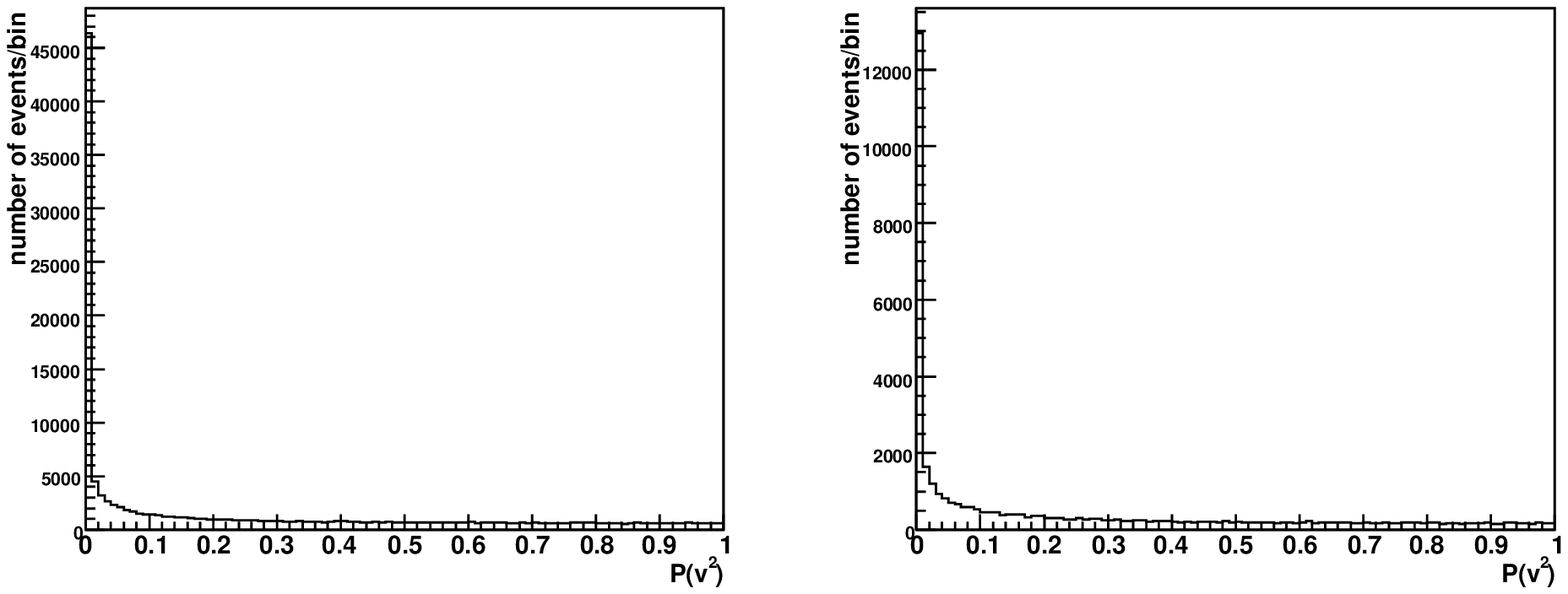}
\caption{\it Distributions of the probability $P(v^2)$ obtained from
non-triggered telescopes for events with 2 non-triggered telescopes
($n_T=2$, left panels) and events with 1~non-triggered telescope
($n_T=3$, right panel). Top panels: gamma-ray showers simulated at a
zenith angle of $18^\circ$, according to a power-law energy spectrum
with a spectral index of 2.2. Bottom panels: hadronic showers as
observed by H.E.S.S.~at a zenith angle of $20^\circ$~in a field of
view with no significant gamma-ray emission.} \label{fig:pchi2g}
\end{center}
\end{figure}

For each zenith angle and for a given value of $n_T$, the average
value $\ln(Q_0)$ and the standard deviation $\sigma_D$ of this
distribution are determined from simulations. This allows to define
the normalised Gaussian variable $v = \ln (Q_{3D}/Q_0)/\sigma_D$,
whose square $v^2$ behaves as a $\chi^2$ with one degree of freedom.
It is thus possible to use the corresponding $\chi^2$ probability
$P(v^2)$ as an additional goodness variable. There is one such
variable per non-triggered telescope which, for gamma-ray showers,
should be uniformly distributed between 0 and 1. This is actually
verified in Fig.~\ref{fig:pchi2g} (top panels), which shows these
distributions for gamma-ray showers simulated at a zenith angle of
$18^\circ$ for $n_T=2$ and $n_T=3$. In the case of $n_T=2$, the
small fraction ($\approx$ 2\%) of gamma-ray events whose predicted
charge $Q_{3D}$ in a non-triggered telescope is less than 20~p.e.
(see Fig.~\ref{fig:qdark}) has been removed from the histogram on
the top left panel of Fig.~\ref{fig:pchi2g}. These events, which
correspond to showers falling at large distance from the centre of
the array, have a very small probability $P(v^2)$ and would have
been rejected by the standard cuts too. There are no such events for
$n_T=3$. Conversely, Fig.~\ref{fig:pchi2g} (bottom panels) shows the
distributions of the probability $P(v^2)$ for hadronic showers as
observed by H.E.S.S.~ at a zenith angle of $20^\circ$~in a field of
view with no significant gamma-ray emission. The accumulation at
very low probabilities is very clear. The dependence of $\ln Q_0$
and that of the standard deviation $\sigma_D$ of the distributions
of $\ln Q_{3D}$ on the telescope multiplicity $n_T$ and on the
zenith angle $\zeta$ are obtained from simulations and are given in
Table~\ref{tab:darkform}. The values found for $Q_0$ are comparable
to typical cutoff values for the charge collected in a given
telescope, as used in other H.E.S.S. analysis methods \cite{hessstd}
\cite{mathieu}.

\subsection{Overall quality criterion and performance of the method}
\label{sec-perf} For any event reconstructed by the 3D-model, 4
goodness variables are available, one for each telescope, whatever
the trigger decision. For gamma-ray showers, these quantities should
be uniformly distributed between 0 and 1, whereas hadronic ones
should most of the time lead to low values. The new quality
criterion requires the goodness variables of all triggered
telescopes to be higher than $p_{\rm  trig} = 0.05$ and the goodness
variables of all non-triggered telescopes to be higher than  $p_{\rm
dark} = 0.07$ . This ensures that the data are compatible with the
basic assumptions of the model (mainly rotational symmetry) and with
the fitted values of the shower parameters. After this first step,
the standard selection cuts described in \cite{3Dmodel} are applied.
They require the compatibility of the fitted parameters with those
of an electromagnetic shower, namely~:
\begin{itemize}
\item compatibility between the depth of shower maximum and the
number of Cherenkov photons (formula (2) of reference
\cite{3Dmodel});
\item restriction on the shower 3D-width, in practice on the
``reduced 3D-width'', whose distribution is almost independent of
the zenith angle (formula (3) of reference \cite{3Dmodel}).
\end{itemize}
We have checked that the distributions of the different goodness
variables defined above did not change significantly after applying
the two preceding standard cuts. Finally, in the case of a
point-like source, a cut is applied on the angle $\theta$ between
the reconstructed direction and that of the source \cite{3Dmodel},
namely $\theta < 0.1^\circ/\cos \zeta$.

\begin{table}[htbp]
\begin{center}
\begin{tabular}{|cc||c|c|c|}\hline
\multicolumn{2}{|c||}{Zenith angle} & $18^\circ$ & $37^\circ$ & $46^\circ$ \\
\hline \hline 3D & $\epsilon_g$ & 0.92 & 0.91 & 0.91 \\
standard & $R_h$ & 9.95 & 9.69 & 10.51 \\
analysis & $Q_{\rm ext}$ & 2.89 & 2.85 & 2.94 \\ \hline \hline
New & $\epsilon_g$ & 0.78 & 0.78 & 0.77 \\
3D & $R_h$ & 21.01 & 18.54 & 18.05 \\
analysis & $Q_{\rm ext}$ & 3.59 & 3.34 & 3.26 \\
 & Gain$_{\, \rm ext}$ & 1.24 & 1.17 & 1.11 \\\hline
\end{tabular}
\caption{\it New 3D analysis as compared to the standard one for an
extended source (no angular cut is applied): $\epsilon_g$ is the
gamma-ray selection efficiency, $R_h$ the hadronic rejection factor
and $Q_{\rm ext} = \epsilon_g \sqrt{R_h}$ the quality factor. The
gain in significance ({\rm Gain}$_{\, \rm ext}$) is given by the
ratio of quality factors.} \label{tab:extended}
\end{center}
\end{table}

\begin{table}[htbp]
\begin{center}
\begin{tabular}{|cc||c|c|c|}\hline
\multicolumn{2}{|c||}{Zenith angle} & $18^\circ$ & $37^\circ$ & $46^\circ$ \\
\hline \hline 3D & $\epsilon_g$ & 0.58 & 0.60 & 0.58 \\
standard & $R_h$ & 5292 & 4233 & 2966 \\
analysis & $Q_{\rm point}$ & 41.9 & 38.71 & 31.82 \\ \hline \hline
New & $\epsilon_g$ & 0.51 & 0.52 & 0.51 \\
3D & $R_h$ & 8468 & 6124 & 4029 \\
analysis & $Q_{\rm point}$ & 47.11 & 40.92 & 32.21 \\
 & Gain$_{\, \rm point}$ & 1.12 & 1.06 & 1.01 \\\hline
\end{tabular}
\caption{\it New 3D analysis as compared to the standard one for a
point-like source (angular cut $\theta < 0.1^\circ/\cos \zeta$):
$\epsilon_g$ is the gamma-ray selection efficiency, $R_h$ the
hadronic rejection factor and $Q_{\rm point} = \epsilon_g
\sqrt{R_h}$ the quality factor. The gain in significance ({\rm
Gain}$_{\, \rm point}$) is given by the ratio of quality factors.}
\label{tab:point}
\end{center}
\end{table}
\begin{table}[htbp]
\begin{center}
\begin{tabular}{|c|c||c|c|c|}\hline
3D standard & Zenith angle & $18^\circ$ & $33^\circ$ & $51^\circ$ \\
\cline{2-5}
 analysis with & $R_h$ (extended)& 16.1 & 13.9 & 11.3 \\
$n_T \ge 3$ & $ R_h$ (point-like)& 6870 & 4980 & 2170 \\ \hline
\end{tabular}
\caption{\it Hadronic rejection factors for extended and point-like
sources in the 3D standard analysis in which the restriction $n_T
\ge 3$ is applied. Results are reproduced from \cite{3Dmodel} which
uses slightly different values of the zenith angle. For the sake of
comparison with Tables \ref{tab:extended} and \ref{tab:point},
linear interpolations in $\cos \zeta$ yield $R_h$ (extended) = 13.7
for $\zeta=37^\circ$ and 12.1 for $\zeta=46^\circ$. Similarly, $R_h$
(point-like) = 4560 for $\zeta=37^\circ$ and 3050 for
$\zeta=46^\circ$.} \label{tab:std}
\end{center}
\end{table}

The new method is compared to the standard 3D
analysis\footnote{Compared to the H.E.S.S. standard analysis, based
on the mean-scaled widths and lengths of shower images, the standard
3D analysis provides a gain of 1.12 in significance \cite{h2356}.}
in Table~\ref{tab:extended} for an extended source (i.e. without any
cut on the shower direction) and in Table~\ref{tab:point} for a
point-like source. For each kind of analysis, gamma-ray selection
efficiencies $\epsilon$ are obtained from simulations assuming a
power-law energy spectrum with photon index 2.2. The loss of genuine
gamma-ray events due to the new criteria ($\sim 15$\% on average) is
almost energy-independent above 100~GeV.
The corresponding rejection factors $R$ for hadrons (i.e. the
factors by which the numbers of hadronic events at the trigger level
are reduced after applying selection cuts) are obtained from real
H.E.S.S. data taken in fields of view with no significant gamma-ray
emission. In the case of a weak source, the significance of the
signal is proportional to the quality factor $Q=\epsilon \sqrt{R}$
which therefore characterises the sensitivity of the method. The
ratio of the quality factors in the new 3D analysis and in the
standard one respectively indicates the gain in significance
obtained from the new method for a source detection. When values of
$p_{\rm trig}$ and $p_{\rm dark}$ lower than those given above are
used, the quality factors and the gain are found to be rather
stable. The values chosen for these two probabilities have the
advantage to significantly reduce the residual contamination of the
selected sample. For an extended source at low zenith angles, the
background is reduced by a factor of 2, which is important for
morphological as well as for spectral studies. The gain in
sensitivity, namely 1.24 at low zenith angles for an extended
source, corresponds to that of an observation time increased by 50\%
in the framework of the standard 3D analysis. For point-like sources
at low zenith angles, the gain is lower albeit still significant and
the background is reduced by a factor 1.6. It should be noted that
the improvement of the method is stronger for events triggering two
telescopes which are the more numerous. The standard 3D analysis had
sometimes been applied with the restriction $n_T \ge 3$ in order to
achieve a higher sensitivity (see e.g. \cite{h2356}). With the new
analysis, the hadronic rejection factors obtained {\it without any
cut on the telescope multiplicity} are greater than or comparable to
those obtained in the standard 3D method with $n_T \ge 3$. From
Table~\ref{tab:std}, in which the corresponding factors are
reproduced from reference \cite{3Dmodel}, it can be verified that
the rejection factors obtained in the new analysis with $n_T \ge 2$
are higher by 20 to 30\%.

\section{Testing the method and the simulations with the almost pure gamma-ray sample from the flaring period of the blazar PKS~2155-304 in July 2006}\label{sec-testing}
 The exceptional VHE gamma-ray flare of PKS
2155-304 during the nights MJD 53944 and MJD 53946, in which the
average intensity of the outburst was about 7 times the flux
observed from the Crab Nebula \cite{pks06}, provided us with an
unprecedented data-set taken under uniform experimental conditions.
The very high quality data featuring an abundance of gamma-rays with
a very small contamination (less than 3\%) of hadronic background
effectively provides us with a gamma-ray test-beam . To correct for
these remaining hadronic events, the reflected background model
subtraction method \cite{berge07} has been applied. In this case,
the OFF region from which the background is estimated is the exact
symmetric counterpart of the ON region with respect to the telescope
axis, so
that acceptances in both regions are the same.\\

The simulations used for the comparison were custom-made to fit the
actual experimental conditions during the observation period, such
as the night sky background rate (40~MHz) and the optical efficiency
of the telescopes, which is directly measured via images of muon
rings \cite{calib}. The latter is important for the calibration of
the energy measurement, which relies on the reconstruction of the
number of Cherenkov photons in the shower and is therefore
particularly sensitive to variations in the relative optical
efficiency.  The data-set has been divided into samples
corresponding to given intervals of zenith angle used in the
simulation and the comparison between data and simulations was
carried out for all these samples. The gamma-ray energy spectrum was
simulated according to a power law, using the spectral index of 3.3
fitted from the data of the two flaring nights.\\

\begin{figure} [htbp]
\begin{center}
  \includegraphics[width=0.45\linewidth]{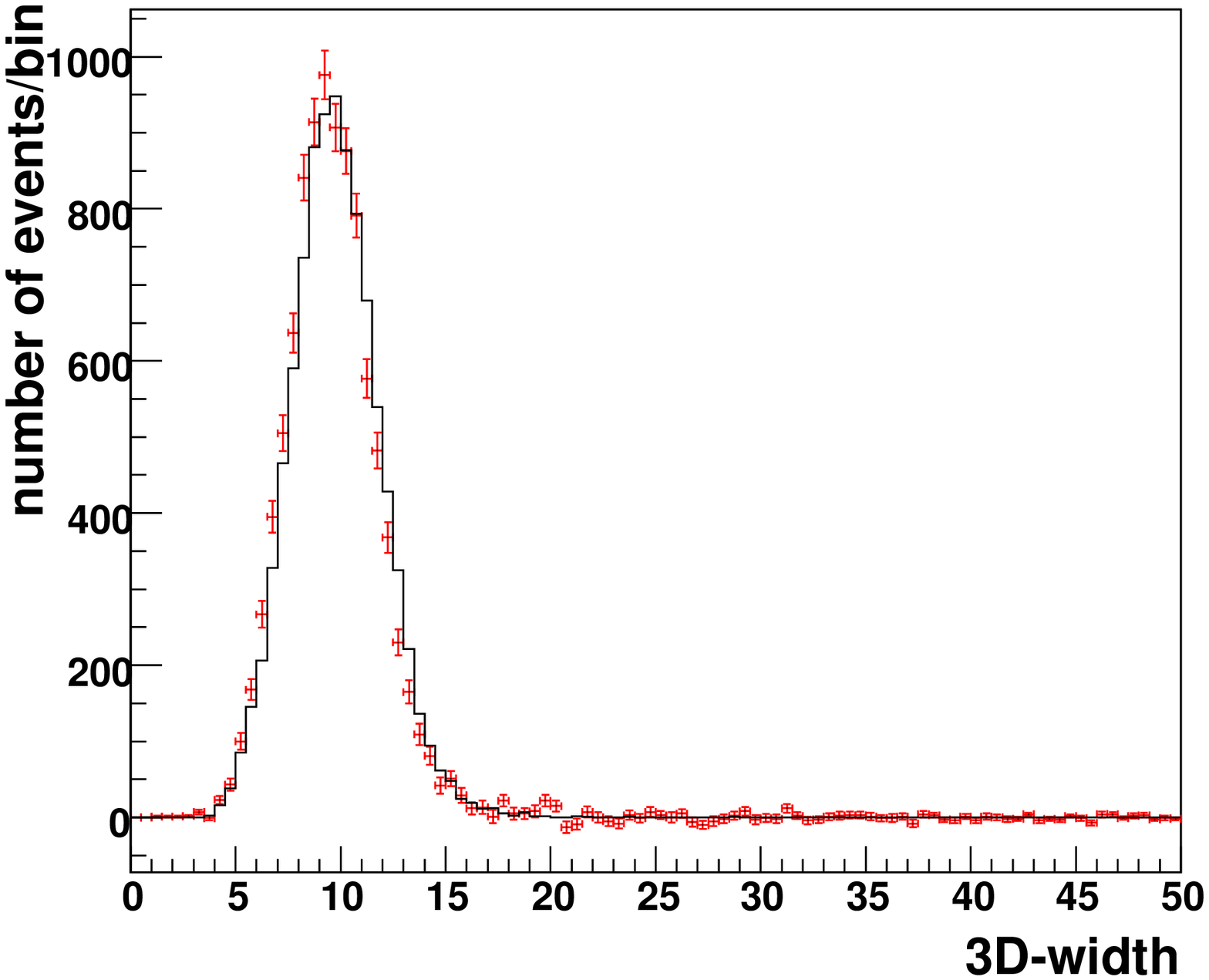}
  \includegraphics[width=0.45\linewidth]{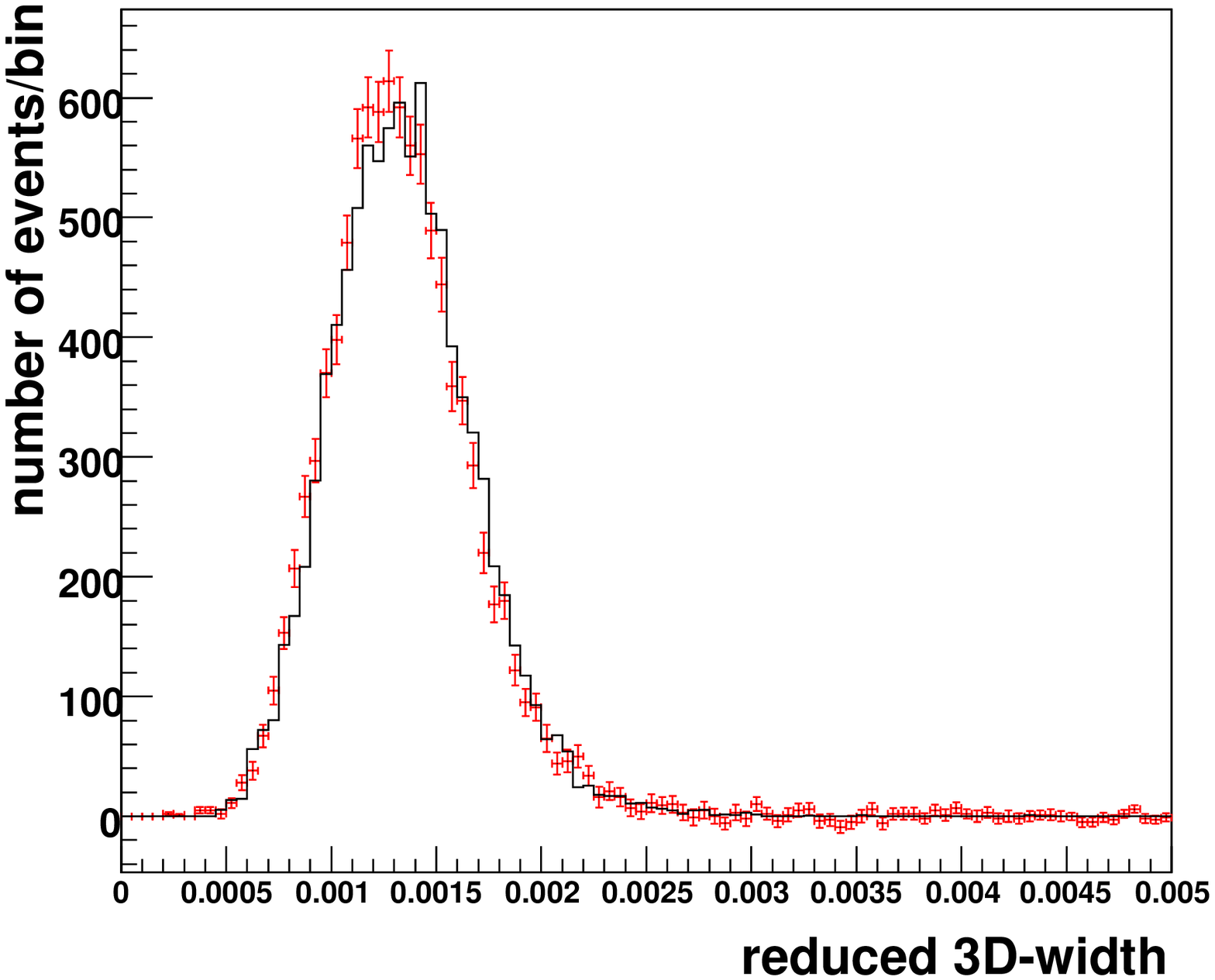}
  \caption{\it Comparison of observable quantities
related to the atmospheric profile between the data (red points with
error bars) and 3D-Model simulations (black histogram). Left panel:
3D-width. Right panel: Reduced 3D-width.}\label{width}
\end{center}
\end{figure}

Observable quantities that are characteristic of the 3D-Model and
related to crucial experimental parameters have been compared
systematically, thereby examining the degree of accuracy of the
match. In the following, all the comparisons refer to a zenith angle
of $18^\circ$ but they are representative for the whole range of
zenith angles covered by this study. For each observable $X$, whose
mean values are $\bar{X}_d$ (real data) and $\bar{X}_s$
(simulations), the relative bias
$2(\bar{X}_d-\bar{X}_s)/(\bar{X}_d+\bar{X}_s)= \Delta
\bar{X}/\bar{X}$ is given in Table \ref{tab:compdis}. All these
quantities are at most of a few percent. The half-widths
half-maximum ($HWHM_d$ for data, $HWHM_s$ for simulations) of all
histograms shown in Fig. \ref{trigger}, \ref{atmo}, \ref{width} and
\ref{NC} (left panel) have been calculated. Table \ref{tab:compdis}
shows that, for all observables, the differences $\Delta (HWHM) =
HWHM_d - HWHM_s$ are smaller than the bin width.
\subsection{Trigger-related quantities}
\begin{figure}
\begin{center}
  \includegraphics[width=0.45\linewidth]{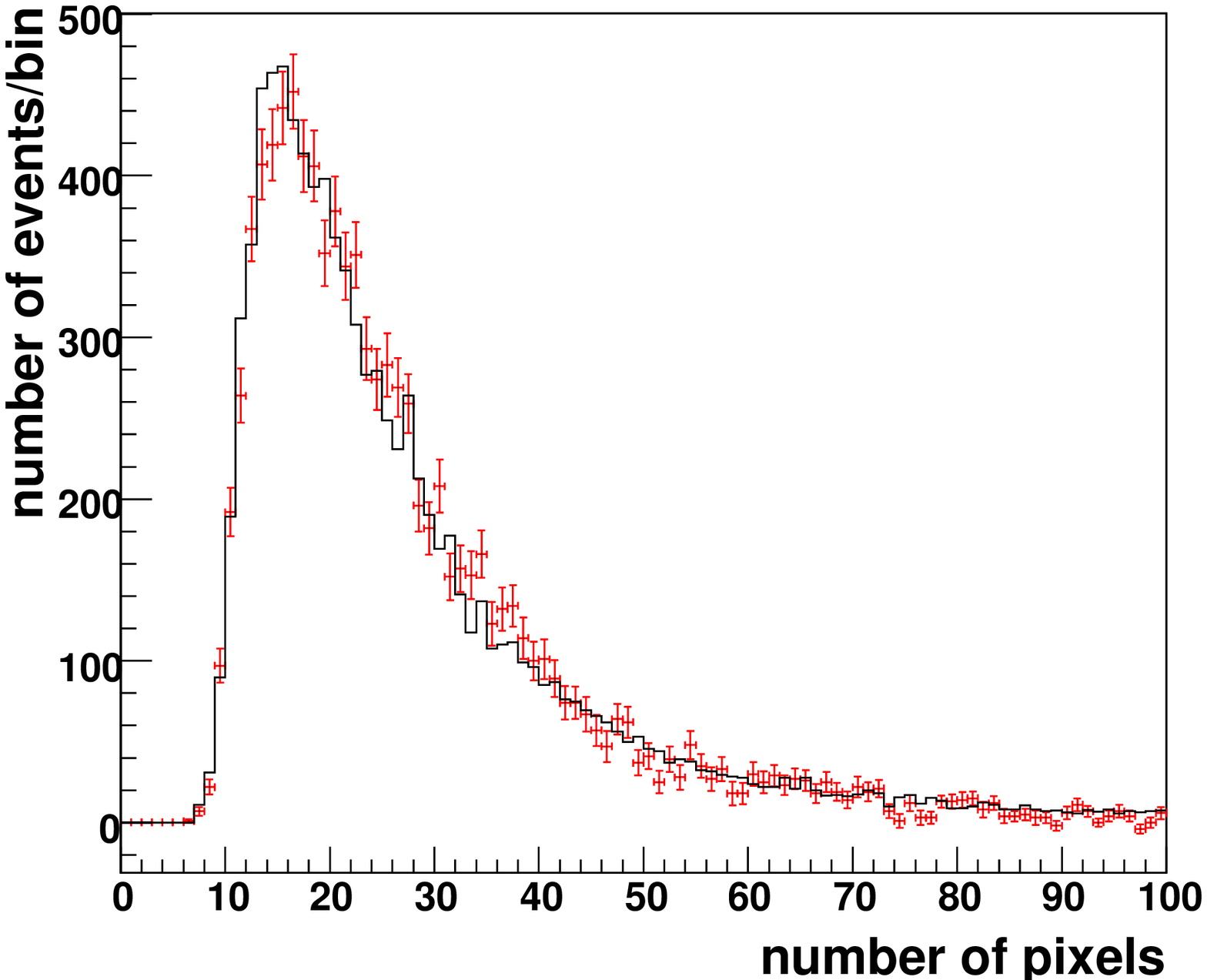}
  \includegraphics[width=0.45\linewidth]{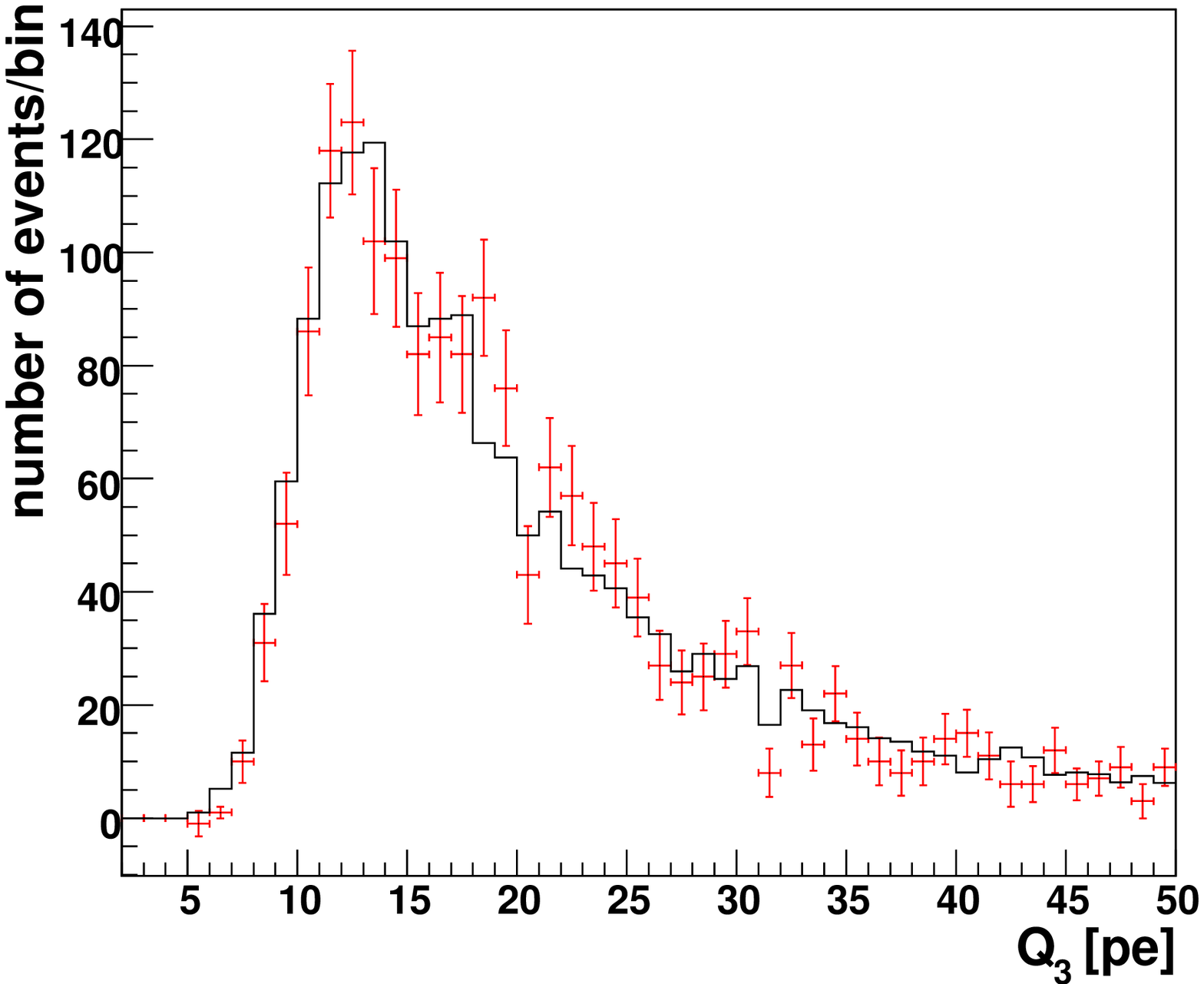}
\caption{\it Comparison of observable quantities related to the
trigger between the data (red points with error bars) and 3D-Model
simulations (black histogram). Left panel: Number of pixels per
telescope. Right panel: Charge in the third most brightest pixel of
one of the telescopes.}\label{trigger}
\end{center}
\end{figure}

Fig.~\ref{trigger} shows the agreement for trigger-related
quantities. The distribution of the number of pixels hit by photons in
the camera is in very good agreement with the simulation. Especially
the rising edge of the distribution, which is particularly sensitive
to the actual experimental conditions, such as the night sky
background rate, has been reproduced by the simulation to a high
degree of accuracy. A second trigger-sensitive variable is the
amount of charge in the third most luminous pixel in each telescope.
As the trigger condition requires a threshold of 4 photo-electrons
in at least three pixels in each sector of the camera, the
comparison of the charge in the third most luminous pixel allows us
to scrutinise the accuracy of the trigger.

\subsection{Atmospheric profile}
\begin{figure} [htbp]
\begin{center}
  \includegraphics[width=0.45\linewidth]{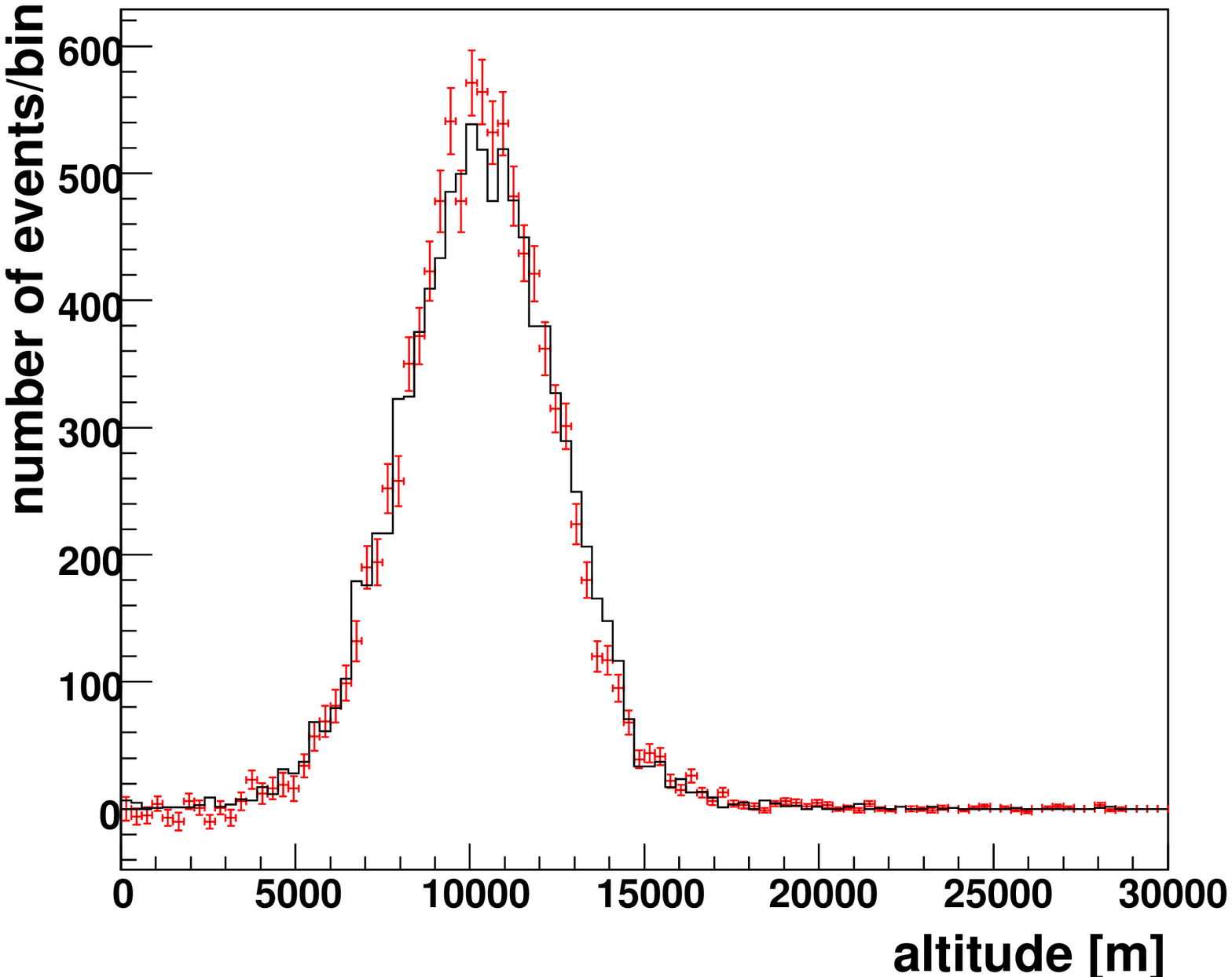}
  \includegraphics[width=0.45\linewidth]{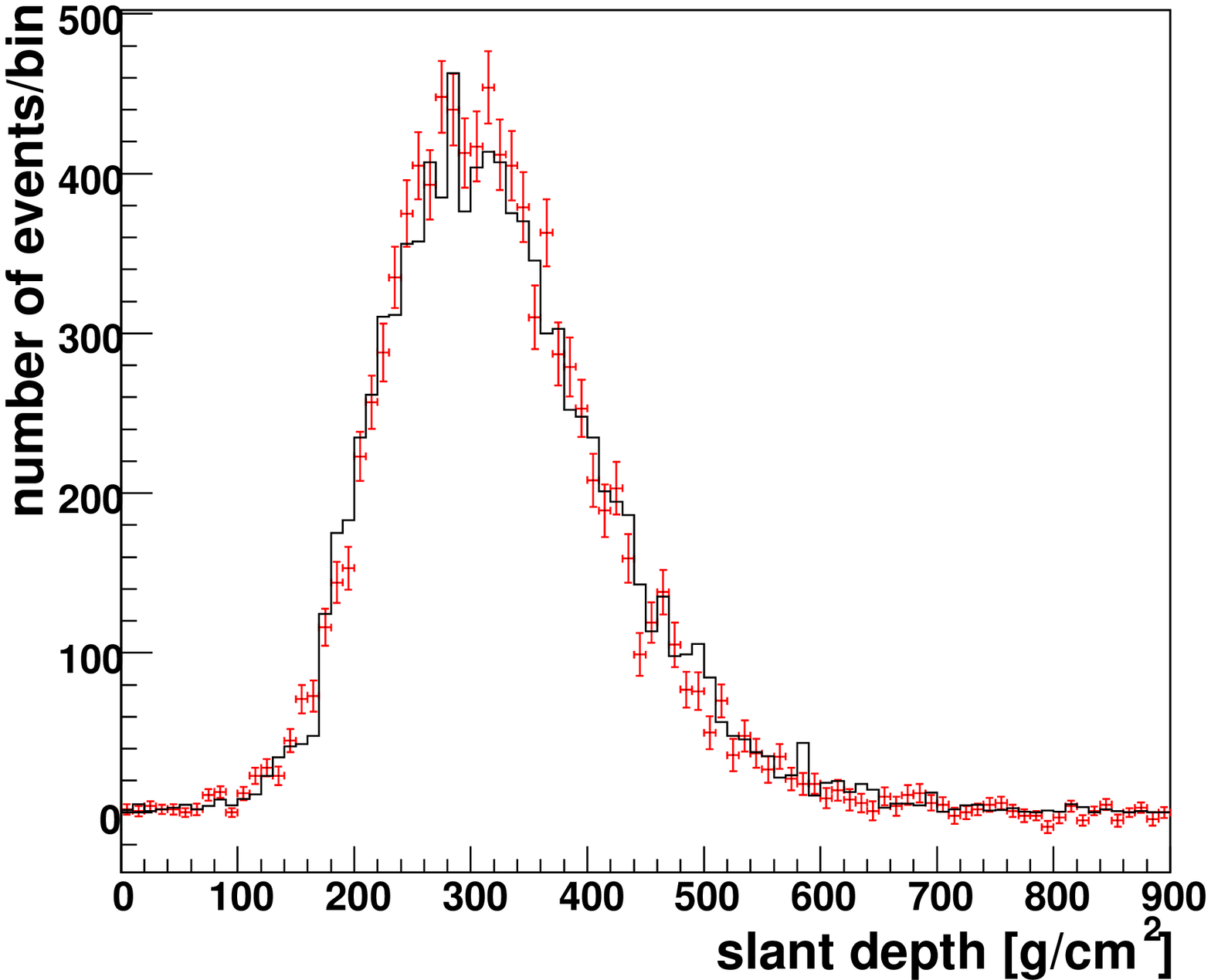}
\caption{\it Comparison of observable quantities related to the
atmospheric profile between the data (red points with error bars)
and 3D-Model simulations (black histogram). Left panel: Altitude of
the shower maximum. Right panel: Slant depth (grammage) of the
shower maximum.}\label{atmo}
\end{center}
\end{figure}

Other environmental conditions studied here include the atmospheric
profile (desert atmosphere) used in the simulation. Whereas the
altitude of shower maximum, as reconstructed by the 3D model, is a
purely geometric parameter, the slant depth of shower maximum
expressed in g/cm$^2$ (grammage) is determined on the basis of the
atmospheric profile. Since the distributions of the first quantity
show a good agreement between data and simulations (see Fig.~\ref{atmo},
left panel), any discrepancy between the true
atmospheric profile and that of simulations would induce a
difference between the distributions of the slant depth obtained
from simulations and data respectively. This is however not observed
in Fig.~\ref{atmo} (right panel), which validates the atmospheric
profile implemented in simulations.

\subsection{Shower characteristics}

Even more pivotal than the agreement of environmental parameters is
the accordance of the observable quantities upon which gamma-rays
are separated from hadronic showers. In the 3D-model, this
separation relies on the reduced 3D-width of the observed shower.
Noting that the real 3D-width of the shower varies significantly
with the zenith angle, a new parameter called the reduced 3D-width
\cite{3Dmodel} was introduced, which is the ratio of the 3D-width to
the slant depth at shower maximum, both being expressed in g/cm$^2$
at the altitude of shower maximum. This last variable was found to
be almost independent of the zenith angle.

\begin{figure} [htbp]
\begin{center}
  \includegraphics[width=0.45\linewidth]{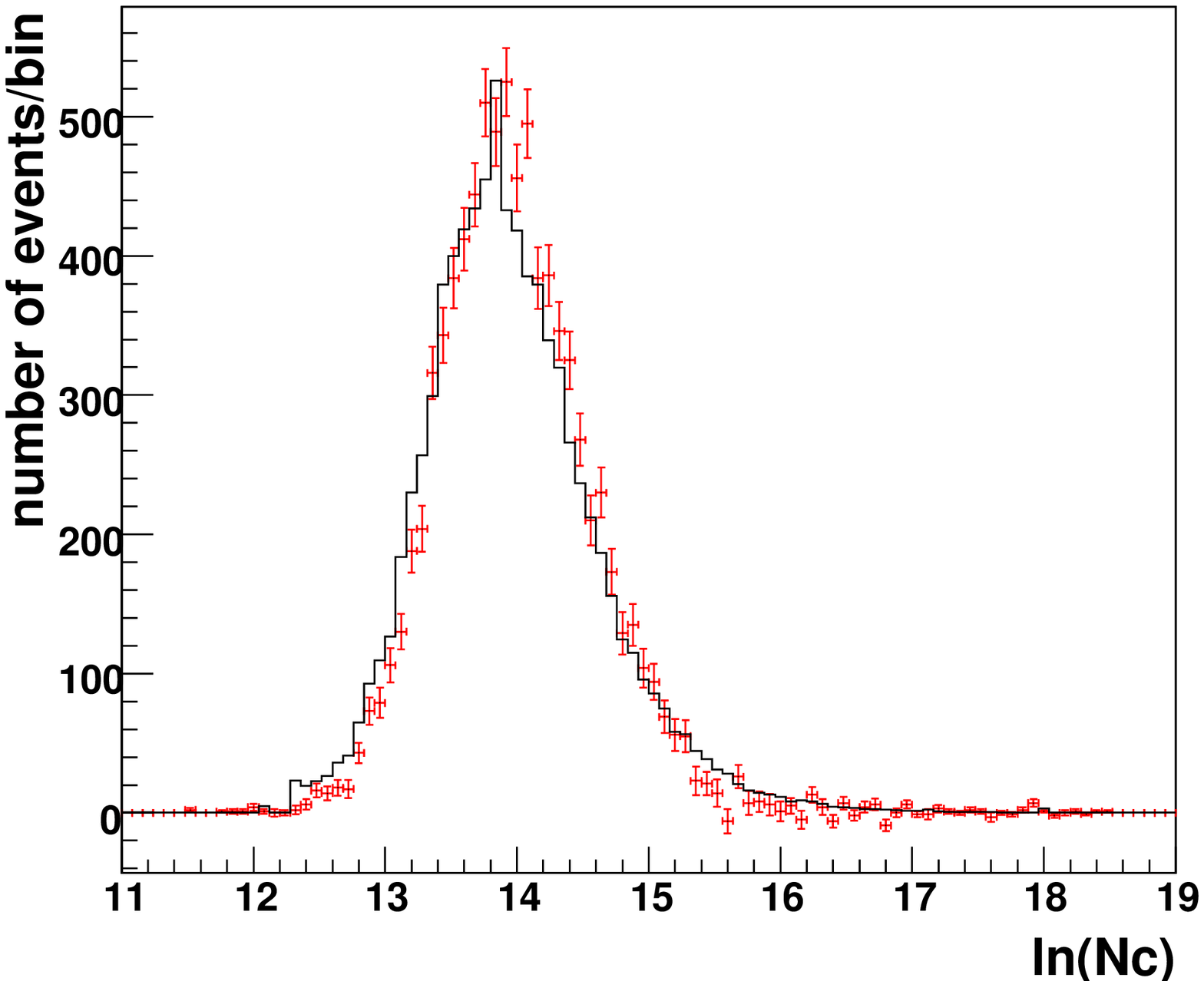}
  \includegraphics[width=0.45\linewidth]{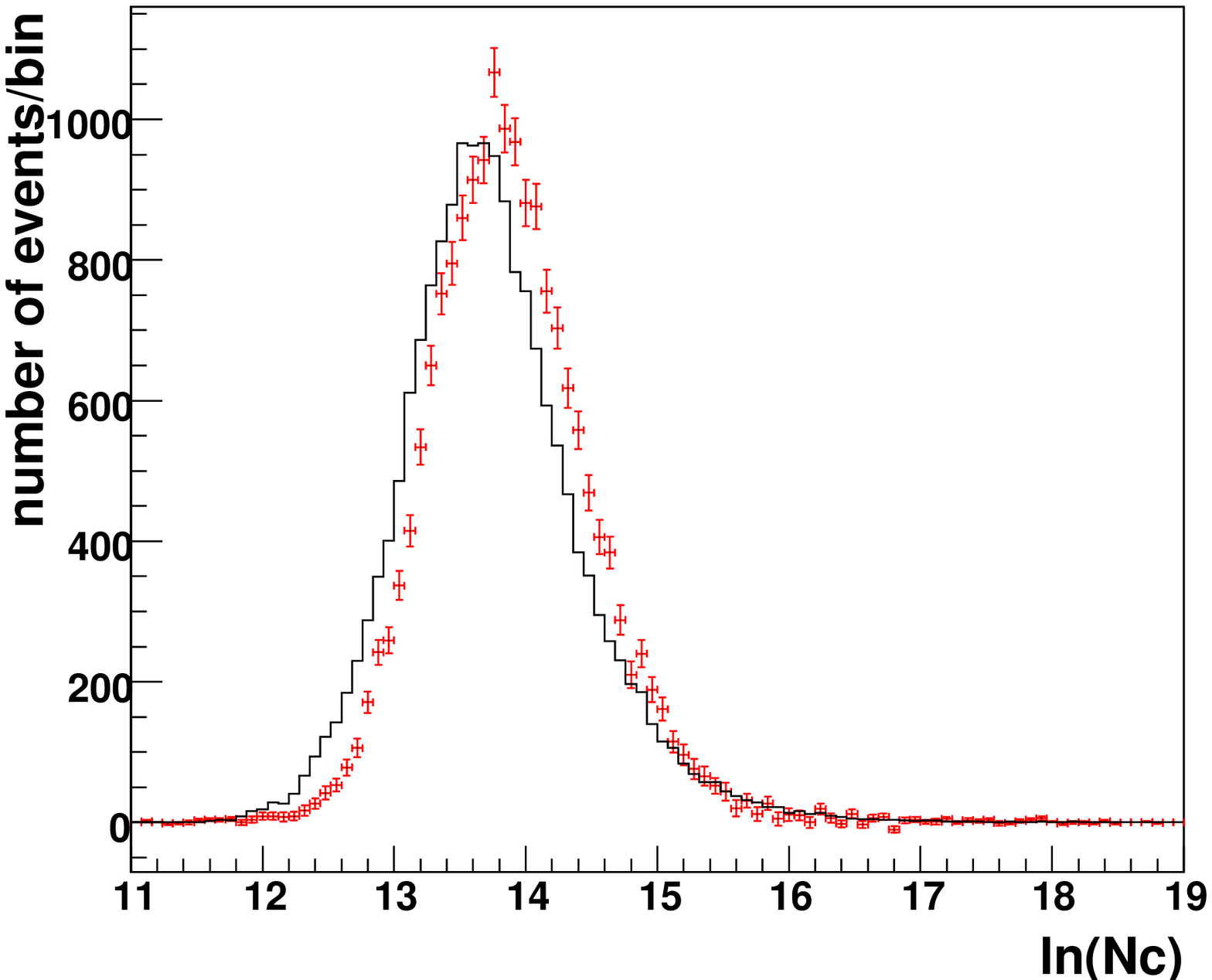}
\caption{\it Comparison of the number of reconstructed Cherenkov
photons for the data (red points with error bars) and 3D-Model
simulations (black histograms). Left panel: the optical efficiencies
have been measured from muon ring images. Right panel: the optical
efficiency has been increased by 10\%.}\label{NC}
\end{center}
\end{figure}

The other crucial observable quantity examined, the logarithm of the
number of Cherenkov photons, is related directly to the energy
measurement. As this is the single parameter in the 3D-model that
has to be calibrated from simulations, its verification is
essential. Fig.~\ref{NC} (left panel) and Table \ref{tab:compdis}
show a good agreement
between data and simulations on the distribution of the logarithm of
the number $N_c$ Cherenkov photons, when the optical efficiency
derived from muon rings \cite{calib} is used in both cases. As a
matter of fact, using the same optical efficiency in simulation and
in data analysis is not sufficient to guarantee a good agreement
between the distributions of $\ln(N_c)$. This is shown in
Fig.~\ref{NC} (right panel), in which this common efficiency has
been increased by 10\% with respect to that given by muon rings
resulting in a manifest shift of the $\ln(N_c)$ distributions.
Therefore, the agreement shown in Fig.~\ref{NC} (left panel) and
Table \ref{tab:compdis} implies that the optical efficiency used in
the simulations (both in the event generation and in its
reconstruction) matches the real one. In contrast to the previously
studied variables, the energy measurement seems to be the one most
affected by variations in the experimental conditions. In
particular, its sensitivity to the relative optical efficiency is
critical and emphasises the need for a careful calibration.
\begin{table}[htbp]
\begin{center}
\begin{tabular}{|c||c|c|}\hline
 & & \\
Observable $X$ & $100 \times \frac{\Delta \bar{X}}{\bar{X}}$ &
$\frac{\Delta ({\rm HWHM})}{\rm Bin \: width}$\\
 & & \\ \hline Number of
pixels per telescope &
-2.82 $\pm$ 3.00 & 0.62 \\
Charge in the 3$^{\rm rd}$ brightest pixel & 2.60 $\pm$ 1.23 & 0.17 \\
Altitude & -0.22 $\pm$ 0.37 & -0.51 \\
Slant depth & -1.11 $\pm$ 0.55 & -0.06 \\
3D-width & -1.53 $\pm$ 0.72 & 0.31 \\
Reduced 3D-width & -0.91 $\pm$ 0.67 & 0.33 \\
$\ln (N_c)$ & 0.18 $\pm$ 0.10 & -0.47 \\ \hline
\end{tabular}
\caption{\it Comparison between simulated and real distributions
shown in Fig. \ref{trigger}, \ref{atmo}, \ref{width} and
 \ref{NC} (left panel). For each observable $X$, whose mean values are $\bar{X}_d$
(real data) and $\bar{X}_s$ (simulations), the relative bias
$2(\bar{X}_d-\bar{X}_s)/(\bar{X}_d+\bar{X}_s)= \Delta
\bar{X}/\bar{X}$ is given in percent units in the second column.
Each distribution has also been characterised by its half-width
half-maximum (HWHM$_d$ for data, HWHM$_s$ for simulations). In the
last columnn, the difference $\Delta (HWHM)$ = HWHM$_d$ - HWHM$_s$
is expressed as a fraction of the bin width.} \label{tab:compdis}
\end{center}
\end{table}
\subsection{Quality criterion, gamma-ray selection efficiency and background rejection}
\begin{figure} [htbp]
\begin{center}
  \includegraphics[width=0.45\linewidth]{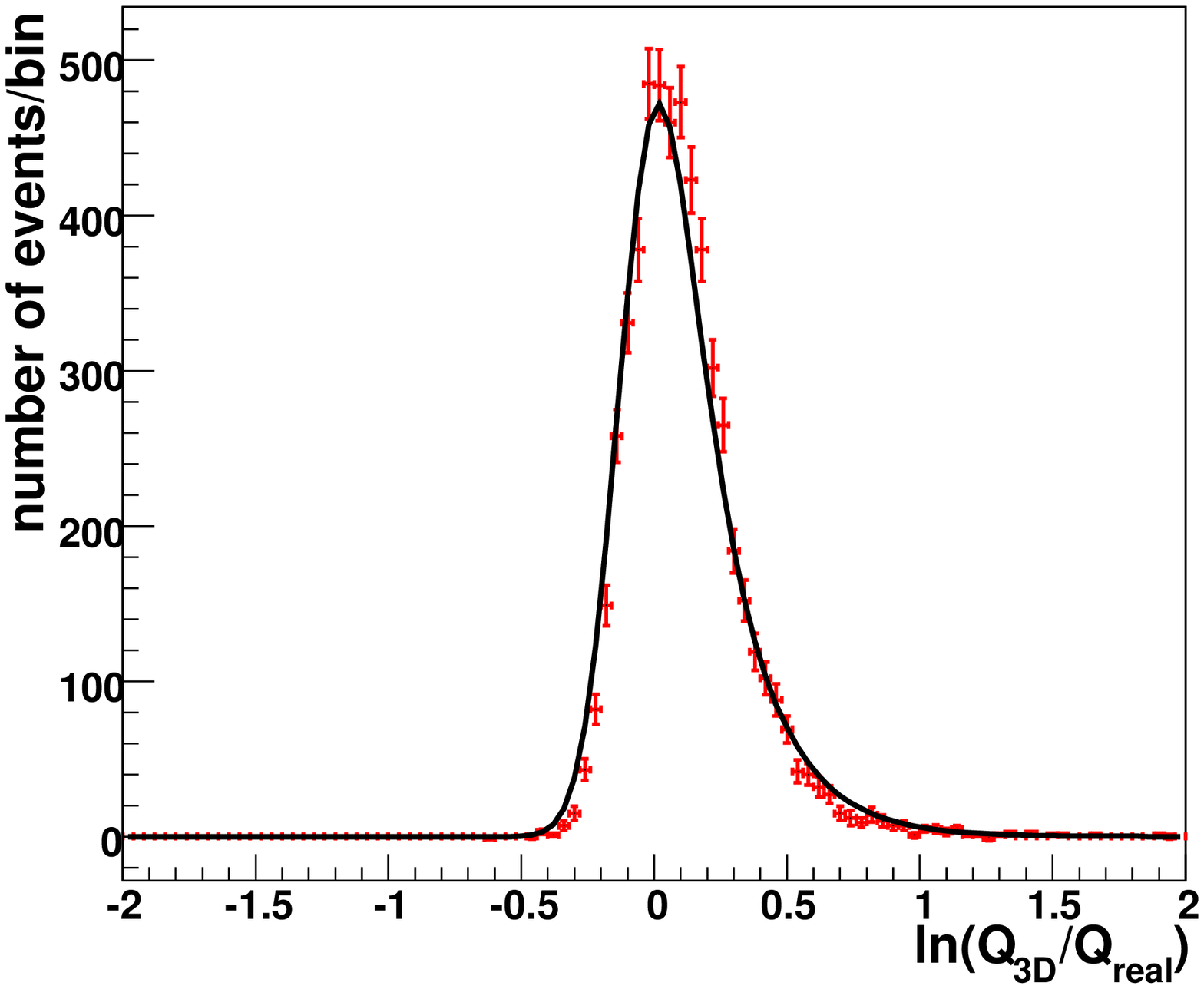}
 \includegraphics[width=0.45\linewidth]{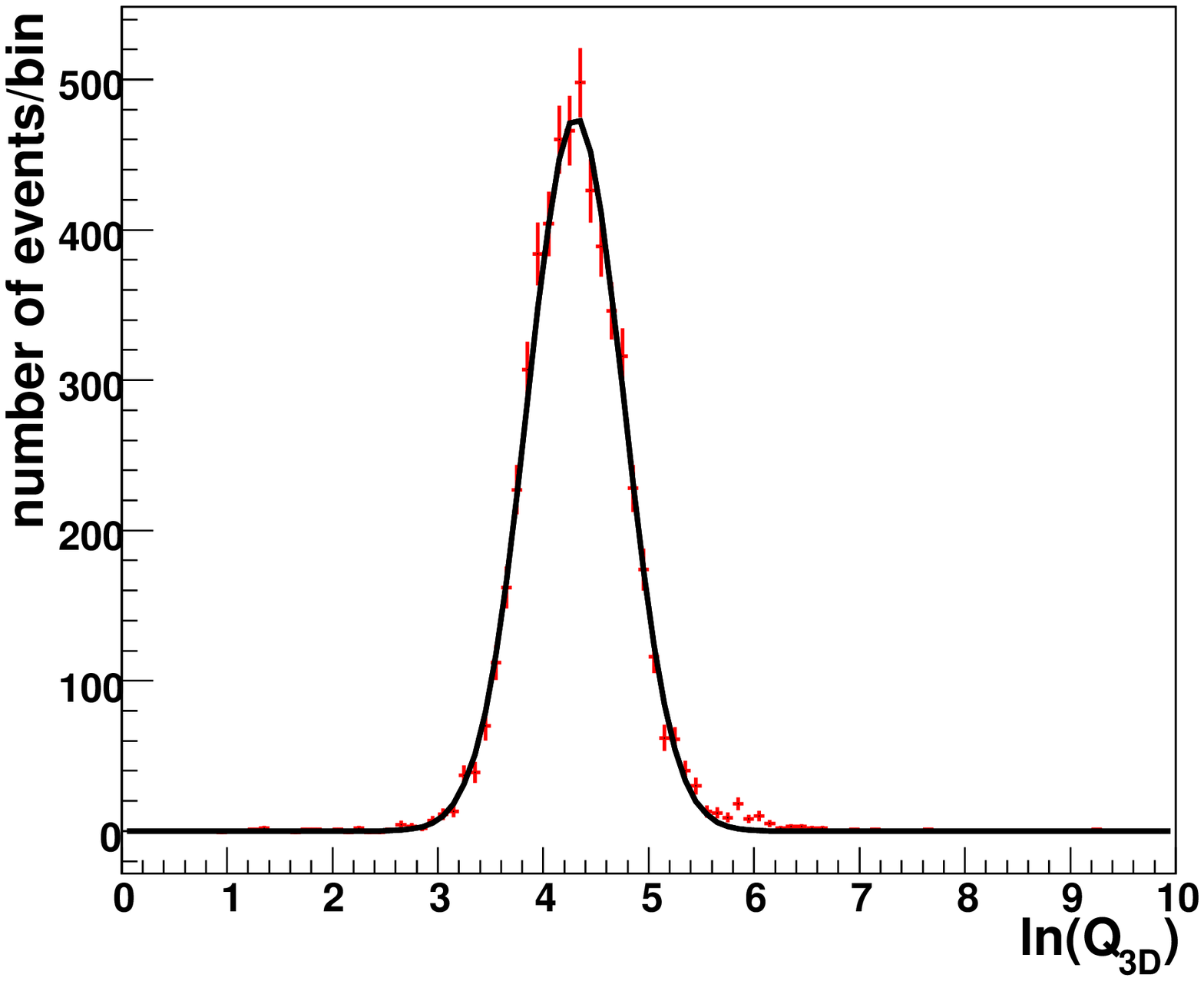}
\caption{\it Left panel: comparison of the distribution of the
variable $u=\ln(Q_{3D}/Q_{\rm real})$ for the data (points with
error bars) for an average zenith angle of 18$^\circ$ and $n_T=2$,
to the curve deduced from formula~(\ref{eq:param}) and
Table~\ref{tab:trigform}. Right panel: comparison of the
distribution of the variable $\ln(Q_{3D})$ for the data (points with
error bars) in the same conditions, to the Gaussian distribution
whose parameters are deduced from Table~\ref{tab:darkform}.}
\label{fig:PKS_gammau2}
\end{center}
\end{figure}

Finally, the performance of the method determined from simulations
and discussed in section~\ref{sec-perf} can be checked on the basis
of real data from the flare of PKS~2155-304. The experimental
distribution of the variable $u=\ln(Q_{3D}/Q_{\rm real})$ defined
for triggered telescopes in section~\ref{sec-qual}, shown by points
with error bars in Fig.~\ref{fig:PKS_gammau2} (left panel) for an
average zenith angle of 18$^\circ$ and $n_T=2$, shows a good
agreement with the curve deduced from formula~(\ref{eq:param}) and
Table~\ref{tab:trigform}. In the right panel of
Fig.~\ref{fig:PKS_gammau2}, the experimental distribution of
$\ln(Q_{3D})$ for non triggered telescopes is found to be in good
agreement with the Gaussian whose parameters are deduced from
Table~\ref{tab:darkform}. It should be noted that both curves were
obtained from simulations based on an energy spectrum with photon
index 2.2, whereas the corresponding index for PKS~2155-304 during
the flaring period is 3.3 \cite{pks06}. This results confirms, on
the basis of real data, that the distributions of the new variables
$u$ and $v$ are not sensitive to the spectral index of the gamma-ray
source under study, as already checked from simulations. The
hadronic rejection factors at 18$^\circ$ zenith angle obtained from
PKS~2155-304 data ($5085 \pm 343$ for the 3D standard analysis and
$7935 \pm 668$ for the new one), are found in good agreement with
the figures given in Table~\ref{tab:point} obtained from simulations
for point-like sources. Only the ratio of gamma-ray selection
efficiencies in the new and in the standard 3D analyses respectively
can be checked experimentally. At 18$^\circ$ zenith angle, this
ratio, found to be 88\% in PKS~2155-304 data, is very close to the
value of 87\% obtained from simulations and derived from
Table~\ref{tab:point}. The preceding agreement between data and
simulations, both on hadronic rejection factors and on gamma-ray
selection efficiencies, validates the estimation of the gain in
sensitivity achieved by the new 3D~analysis given by
Tables~\ref{tab:extended} and \ref{tab:point}.

\section{Conclusion}
The reconstruction method based on the 3D model of electromagnetic
showers has been improved by an additional consistency check:
namely, the modelled Cherenkov photosphere must reasonably reproduce
the global light yield of all triggered telescopes and be compatible
with a low amount of light for non-triggered ones. The new criteria
involved in this consistency check are quasi-insensitive to the
energy spectrum of gamma-rays used in the simulations. This was
already the case for the gamma-ray/hadron discrimination criteria
described in the 3D standard analysis \cite{3Dmodel} and essentially
based on well-known properties of electromagnetic showers:
rotational symmetry and lateral spread of the Cherenkov photosphere
at shower maximum. This makes this method quite different from those
based on machine learning algorithms or training procedures (e.g.
\cite{tmva}) whose sensitivity is optimised for a given gamma-ray
spectrum. The gain in sensitivity with respect to the 3D standard
analysis corresponds to that of an observation time increased by
50\% for an extended source at low zenith angles, the hadronic
background being lowered by a factor of 2. This contamination is
particularly reduced for the sample of events triggering only two
telescopes, i.e. for the majority of events. Finally, the almost
pure gamma-ray beam provided by the exceptional flare of the blazar
PKS~2155-304 in July 2006 allowed to check the simulation of the
H.E.S.S. experiment at the level of a few percent and to validate
the gamma-ray energy measurement based on the 3D analysis. The
present method, which takes account of the correlations between
images observed by different telescopes and allows a rather fast
processing of the events, will be particularly well suited to the
exploitation of forthcoming large Cherenkov arrays.

\section{Acknowledgments}
We thank Prof.~W.~Hofmann, spokesman of the H.E.S.S.~Collaboration
and Prof.~G.~Fontaine, chairman of the Collaboration board, for
allowing us to use H.E.S.S.~data in this publication. We are
grateful to Dr.~M.~de Naurois for carefully reading the manuscript
and for providing us with very useful suggestions. Finally, our
thanks go to all the members of the H.E.S.S.~Collaboration for their
technical support and for many stimulating discussions.





\begin{thebibliography}{200}
\bibitem{punch} M. Punch, in ``Towards a Network of Atmospheric Cherenkov Detectors
VII'', Palaiseau, France (2005), p. 379, B.~Degrange et G.~Fontaine
ed.
\bibitem{cangaroo} M. Mori, in ``Towards a Network of Atmospheric Cherenkov Detectors
VII'', Palaiseau, France (2005), p. 19, B.~Degrange et G.~Fontaine
ed.
\bibitem{hess} K. Bernl{\"o}hr, et al.,  Astropart. Phys. {\bf 20}
(2003) 111.
\bibitem{veritas} T.C. Weekes, in ``Towards a Network of Atmospheric Cherenkov
Detectors VII'', Palaiseau, France (2005), p. 3, B.~Degrange et
G.~Fontaine ed.
\bibitem{magic} M. Teshima, in ``Towards a Network of Atmospheric Cherenkov
Detectors VII'', Palaiseau, France (2005), p. 373, B.~Degrange et
G.~Fontaine ed.
\bibitem{3Dmodel} M. Lemoine-Goumard, B. Degrange and M.~Tluczykont,
Astropart. Phys., {\bf 25} (2006) 195.
\bibitem{trig} S. Funk, et al., Astropart. Phys. {\bf 22} (2004)
285.
\bibitem{berge07} D. Berge, S. Funk and J. Hinton,
A\&A {\bf 466} (2007) 1219.

\bibitem{h2356} F. Aharonian, et al., A\&A {\bf 455} (2006) 461.
\bibitem{velajr} F. Aharonian, et al., ApJ {\bf 661} (2007) 236.
\bibitem{vacanti} G. Vacanti, et al., Astropart. Phys. {\bf 2}
(1994) 1.
\bibitem{calib} F. Aharonian, et al., Astropart. Phys. {\bf 22}
(2004) 109.
\bibitem{pks06} F. Aharonian, et al., ApJ {\bf 664} (2007) L71.
\bibitem{kascade} J. Guy, th{\`e}se de doctorat, Universit{\'e} Paris VI
(2003); this program is a modified version of that described in
M.~P.~Kertzman and G.~H.~Sembroski, Nucl. Instr. Meth. Phys. Res. A,
{\bf 343} (1994) 629.
\bibitem{smash} J. Guy, th{\`e}se de doctorat, Universit{\'e} Paris VI (2003).
\bibitem{hessstd} F. Aharonian, et al., A\&A {\bf 457} (2006) 899.
\bibitem{gaisser} T.K. Gaisser, in Cosmic Rays and Particle Physics,
Cambridge University Press (1990), p.245.
\bibitem{mathieu} M. de Naurois, in ``Towards a Network of Atmospheric Cherenkov
Detectors VII'', Palaiseau, France (2005), p. 149, B.~Degrange et
G.~Fontaine ed.
\bibitem{tmva} S. Ohm, C. van Eldik and K. Egberts, Astropart. Phys. submitted.\\
F.~Dubois, G.~Lamanna and A.~Jacholkowska, Astropart. Phys. submitted.
\end{thebibliography}
\end{document}